\documentclass[useAMS,usenatbib,usedcolumn]{mn2e}

\newcommand{\ltsimeq}{\raisebox{-0.6ex}{$\,\stackrel
        {\raisebox{-.2ex}{$\textstyle <$}}{\sim}\,$}}
\newcommand{\gtsimeq}{\raisebox{-0.6ex}{$\,\stackrel
        {\raisebox{-.2ex}{$\textstyle >$}}{\sim}\,$}}

\begin{document}

\title[Implications from submm follow-up of radio samples]{Implications for
 unified schemes from submillimetre and far-infrared follow-up of radio-selected samples}
\author[J.A. Grimes, S. Rawlings \& C.J.Willott]
  {Jennifer A.~Grimes$^{1}$, 
  Steve Rawlings$^{1\star}$ and 
  Chris J.~Willott$^{1,2}$ \\ 
  $^1$University ~of ~Oxford, Astrophysics, Keble ~Road, Oxford, OX1 ~3RH, UK \\
  $^2$Herzberg Institute of Astrophysics, National Research Council, 5071 West 
Saanich Road, Victoria, BC, V9E 2E7, Canada\\}




\maketitle

\begin{abstract}

\noindent
We extend our previous analysis which used generalized luminosity
functions (GLFs) to predict the number of quasars and galaxies in
low-radio-frequency-selected samples as a function of redshift, 
radio luminosity, narrow-emission-line luminosity and type of unified scheme \citep[][]{grw2003}. 
Our extended analysis incorporates the observed submillimetre ($850~ \mu$m) flux 
densities of radio sources, employs a new method which allows us to
deal with non detections, and focuses on the high-luminosity population.
First, we conclude that the submillimetre luminosity $L_{850}$ of 
low-frequency-selected radio sources is correlated with the bolometric luminosity $L_{\rm Bol}$ of 
their quasar nuclei via an approximate scaling relation $L_{850} \propto L_{\rm Bol}^{0.7 \pm 0.2}$. 
Second, we conclude that there is quantitative evidence for a receding-torus-like 
physical process for the high-luminosity population within a  
two-population unified scheme for radio sources; this evidence comes from the fact that 
radio quasars are brighter in both narrow emission lines and submillimetre luminosity than 
radio galaxies matched in radio luminosity and redshift. 
Third, we note that the combination of a receding-torus-like scheme 
and the assumption that the observed submillimetre emission is dominated by quasar-heated 
dust yields a scaling relation $L_{850} \propto L_{\rm Bol}^{0.5}$ which is within the errors of that 
determined here for radio-selected quasars, and consistent with that inferred for  
radio-quiet quasars by \citet{wrg}. 

\end{abstract}


\footnotetext{$^{\star}$Contact email: sr@astro.ox.ac.uk}

\section{Introduction}
\label{sec:intro}

Unified schemes for radio galaxies and radio quasars propose that both types of objects are
physically similar but viewed at different angles between the
radio axis and the line of sight. A comprehensive introduction to the subtleties 
involved in such schemes can be found in \citet[][hereafter GRW]{grw2003} in which we presented a method 
of using generalized luminosity functions (GLFs) to predict the number of radio galaxies
and radio-loud quasars as a function of redshift $z$, 
radio luminosity, narrow-emission-line luminosity and type of unified scheme. 
The GRW analysis introduced two new parameters: $\alpha$ 
encoding the well known, and strong, correlation between radio and narrow-line luminosity; 
and $\beta$ encoding the scatter about this relation. GRW concluded that
the simplest unified schemes, in which the probability of a given object being viewed 
as a quasar is independent of $\alpha$, are strongly ruled out by the data 
available on radio-selected samples. However, their
analysis was unable to discriminate between the two most popular modifications to the 
simplest schemes, namely: (i) models in which the probability of a given object being 
viewed as quasar is dependent on $\alpha$ and $\beta$, as in the
`receding torus' scheme \citep{law,simp}; and (ii) models in which there are 
two populations of radio sources, only one of which necessarily comprises of an admixture of
unobscured and obscured quasar nuclei \citep{wqf}. Both modified schemes were able to explain 
both the systematic increase in the fraction of quasars with $\alpha$
and the systematically higher narrow-emission-line luminosities for radio quasars over
radio galaxies, matched in radio luminosity and redshift.

Differences in luminosities between samples of radio quasars and radio galaxies, 
matched in radio luminosity and redshift, have also been found  
at submillimetre and infrared wavelengths \citep[e.g.][]{wsubmm}.
These wavebands would seem to be ideal
for testing unified schemes. The putative opaque dust torus is a vital
ingredient of all unified schemes as it is needed to hide the active nucleus by
absorbing the soft X-ray, UV, optical and near-infrared radiation.
Much of this absorbed energy will be re-radiated by dust grains
at infrared wavelengths, and for high-redshift objects this may be observed
in the submillimetre waveband. This dust is expected
to be optically thin (at least at longer wavelengths) and its thermal emission is
radiated isotropically \citep{efrr94}.

\citet{wsubmm} found that radio-loud quasars were a factor of approximately
$2-5$ times brighter than radio galaxies from SCUBA data at $850 ~ \mu$m, where the 
large uncertainty arises because of the difficulties involved with correcting the
submillimetre flux densities for synchrotron contamination.
However, earlier studies of far-infrared differences between
quasars and radio galaxies led to conflicting results. 
Broad-lined objects were found to be more luminous than narrow-lined
objects in the far-infrared from four pairs of objects in the redshift
range $0.25 < z < 0.57$ \citep{vanb}, whereas \citet{meisenheimer} found no difference
in the mid- and far-infrared flux densities of ten pairs of radio galaxies
and radio quasars with $z \le 2$. Such results have led to discussions 
of the viability of unified schemes for radio-selected quasars and galaxies.

In this paper, submillimetre data on radio-selected samples 
will be incorporated into the GLF approach of GRW. 
The availability of submillimetre data for complete samples of radio sources is much
more limited than is the case for narrow emission-line data, and non detections
are far more common. However, the available submillimetre data has been collected 
in a way that yields quantifiable biases.
We have four main aims. First, to discriminate robustly between
the competing unified models that were considered in GRW. Second, 
to quantify the correlation between $\alpha$ and submillimetre (850 $\mu$m)
luminosity $L_{850}$, leading to an estimate of the correlation between (the unobscured) 
optical quasar luminosity $M_B$ (or, equivalently, bolometric quasar luminosity 
$L_{\rm Bol}$) of radio sources and $L_{850}$. Third, to obtain increased predictive power 
for the GLFs due to the addition of submillimetre data. 
Fourth, to understand the discrepancies between
previous studies of differences between radio galaxies and quasars in
the far-infrared and submillimetre. 

This paper is organized as follows. In
Sec.~\ref{sec:data}, the data to be used are described.
In Sec.~\ref{sec:ch3Lacor}, the correlations between
submillimetre luminosity and the principal components
found in GRW are investigated.
In Sec.~\ref{sec:ch3glf}, new GLFs based on this correlation are 
described and the
results and analysis are presented. We make comparisons between 
the predictions of our GLFs and previous studies in Sec.~\ref{sec:ch3comp}.
In Sec.~\ref{sec:scaling} we derive a scaling relation between far-infrared luminosity and the
bolometric luminosity of the quasar nuclei of radio sources.
We assume throughout that
${\rm H_0} = 70 ~ {\rm km\,s^{-1}\,Mpc^{-1}}, \Omega_{\rm M} = 0.3$ and 
$\Omega_{\Lambda} = 0.7$

\section{Data}
\label{sec:data}

A sample of eighteen high-redshift ($1.3 < z < 3.5$) radio quasars with
existing $850$-$\mu$m SCUBA data has been defined. These are drawn from a suite
of low-frequency-selected redshift surveys: the 3CRR sample \citep{lrl}, the 3CR sample \citep{spinrad};
the 6CE sample \citep{rel01}; parts I and II of the 7CRS sample \citep{w02}; and
the TOOT (T{\small EX}O{\small X}-1000) Survey \citep{hr2003}. The first four of these
surveys have complete optical identifications and essentialy complete spectroscopic redshifts
(or, in the case of the 7CRS, reliable photometric redshifts for the few objects without secure
spectroscopic values). The subsections of the TOOT survey used in this work have 
complete spectroscopy but significantly lower redshift completeness \citep{hr2003}: we are 
confident, however, that few, if any, quasars with $R \le 23.5$ have been overlooked.

\citet{wsubmm} presented SCUBA data for eleven quasars in the redshift range 
$1.37 < z < 2.0$: two quasars from the 6CE sample with 
$2.0 \le S_{151}  \le 3.93$ Jy (6C~0955+3844 and 6C~1045+3513); seven quasars
from the 3CRR sample, within the Right Ascension (RA) 
range $04^{\rm h} < \rm{RA} < 22^{\rm h}$; and two quasars (3C~298 and 3C~280.1)
from the 3CR catalogue, which were not included in the 3CRR for reasons of
low declination and (slightly) low radio flux-density respectively. 
The sky area for the combined 3CRR/3CR subsample was estimated by 
taking $\frac{9}{10}$ of the sky area that the 3CR survey would have with the 
reduced RA range, since all of the 3CRR objects are included in the 3CR survey, but 
one 3CR object (3C~418) is in the correct redshift and RA range but could not be considered 
here because it has no SCUBA data.

\citet{rawlings} presented submillimetre data for eight quasars with $2.5 \le z \le 3.5$:
five radio quasars selected from the TOOT00 and TOOT08 sub-regions of the 
TOOT with $R \le 23.5$ and $0.1 \le S_{151}  < 0.2$ Jy,
where $S_{151}$ is the 151-MHz flux density; and three radio
quasars from parts I and II of the 7CRS sample with $S_{151}  > 0.5$ Jy.
We have halved the effective sky area of the 7CRS survey to account for the fact that 
only two out of four 7CRS quasars in the redshift range are used; this assumes that the
excluded objects, synchrotron-dominated quasars (5C~6.288 and 5C~6.291; one of which was observed by
Rawlings et al.), are unbiased with respect to their (isotropic) thermal dust 
luminosities. 

The [OIII] luminosities of the TOOT objects were estimated from the Lyman-$\alpha$ line 
luminosities (which are dominated by emission from the narrow-line region) from
\citet{raw2} and converted to $L_{\rm [OIII]}$ using the line ratios 
presented in \citet{mc}. [There are uncertainties in this procedure because reddening
can have a large impact on the Ly$\alpha$ / [OIII] ratio.]

For the two 7CRS objects, the [OIII] line luminosity was estimated by assuming a 
rest-frame equivalent width of $30 ~ \rm \AA$ for this line. For the 6CE and 3CR objects,
the line luminosities are taken from \citet{wsubmm}, except 3C437 which
had its [OIII] luminosity estimated using the [OIII]/[OII] relation from GRW,
and 6C~0919+3806 which now correctly employs the CII]/[OIII]
ratio of \citet{mc}.

\begin{table*}
\begin{center}
\small
\begin{tabular}{lccr@{.}lccr@{ $\pm$ }rr@{ $\pm$ }rcrr}
\hline\hline
\multicolumn{1}{l}{ Name}&\multicolumn{1}{c}{Class}&\multicolumn{1}{c}{Redshift}&\multicolumn{2}{c}{$D /$}&\multicolumn{1}{c}{$\log_{10}[L_{151} /$}&\multicolumn{1}{c}{$\log_{10}[L_{\rm [OIII]} /$}&\multicolumn{2}{c}{$S_{850} /$}&\multicolumn{2}{c}{$S_{450} /$}&\multicolumn{1}{c}{SC}&\multicolumn{1}{c}{$\alpha$}&\multicolumn{1}{c}{$\beta$}\\
&&$z$&\multicolumn{2}{c}{kpc}&\multicolumn{1}{c}{$\rm W Hz^{-1} sr^{-1}]$}&\multicolumn{1}{c}{W]}&\multicolumn{2}{c}{mJy}&\multicolumn{2}{c}{mJy}&\multicolumn{1}{c}{}&&\\
\hline\hline
TOOT00$\textunderscore$1214&Q&$3.084$&$\simeq38$&$0^{\dagger}$&$26.77$&$36.25$&$2.04$&$1.18$&$3.80$&$11.10$&$0.00$&$0.030$&$-0.027$\\
TOOT00$\textunderscore$1261&Q&$2.544$&$\simeq24$&$0^{\dagger}$&$26.77$&$36.28$&$1.98$&$1.31$&$-18.90$&$22.80$&$0.00$&$0.031$&$-0.029$\\
TOOT08$\textunderscore$061&Q&$3.277$&$<41$&$0$&$27.06$&$36.89$&$\mathbf{8.59}$&$1.10$&$\mathbf{14.90}$&$5.70$&$0.00$&$0.065$&$-0.040$\\
TOOT08$\textunderscore$079&Q&$3.002$&$<42$&$0$&$26.91$&$36.98$&$0.05$&$1.14$&$-2.50$&$9.10$&$0.00$&$0.063$&$-0.049$\\
TOOT08$\textunderscore$094&Q&$3.256$&$<42$&$0$&$27.20$&$35.82$&$0.91$&$1.05$&$-5.10$&$6.50$&$0.00$&$0.031$&$0.005$\\
5C~6.95&Q&$2.877$&$119$&$3$&$27.55$&$36.41$&$2.40$&$1.23$&$0.30$&$13.40$&$0.00$&$0.067$&$-0.004$\\
5C~7.70&Q&$2.617$&$14$&$4$&$27.75$&$35.35$&$0.32$&$1.20$&$5.90$&$14.80$&$0.00$&$0.035$&$0.043$\\
3C~181&Q&$1.382$&$48$&$0$&$28.11$&$37.37$&$\mathbf{5.27}$&$1.06$&$-0.80$&$7.50$&$0.06$&$0.124$&$-0.018$\\
3C~191&Q&$1.952$&$41$&$1$&$28.52$&$37.21$&$\mathbf{6.39}$&$1.06$&$\mathbf{28.90}$&$8.10$&$0.30$&$0.133$&$0.005$\\
3C~205&Q&$1.534$&$152$&$4$&$28.19$&$37.00$&$\mathbf{2.36}$&$1.10$&$11.60$&$7.90$&$0.52$&$0.113$&$-0.001$\\
3C~268.4&Q&$1.400$&$91$&$9$&$27.97$&$37.25$&$\mathbf{5.13}$&$1.18$&$9.90$&$12.50$&$0.58$&$0.114$&$-0.018$\\
3C~270.1&Q&$1.519$&$101$&$6$&$28.24$&$36.51$&$\mathbf{7.42}$&$1.17$&$15.40$&$11.50$&$0.40$&$0.097$&$0.020$\\
3C~280.1&Q&$1.659$&$39$&$8$&$28.36$&$36.72$&$\mathbf{5.10}$&$1.74$&$0.40$&$23.80$&$0.35$&$0.109$&$0.016$\\
3C~298&Q&$1.439$&$12$&$7$&$28.57$&$37.26$&$\mathbf{21.13}$&$2.06$&$-8.00$&$23.70$&$0.38$&$0.137$&$0.005$\\
3C~318&Q&$1.574$&$6$&$8$&$28.11$&$37.18$&$\mathbf{7.78}$&$1.00$&$\mathbf{21.60}$&$10.60$&$0.00$&$0.116$&$-0.010$\\
3C~432&Q&$1.785$&$109$&$8$&$28.32$&$36.68$&$\mathbf{7.93}$&$1.70$&$2.91$&$20.90$&$0.05$&$0.106$&$0.016$\\
6C~0995+3844&Q&$1.405$&$181$&$3$&$27.45$&$36.58$&$-0.01$&$1.11$&$8.90$&$7.80$&$0.00$&$0.069$&$-0.014$\\
6C~1045+3513&Q&$1.594$&$1$&$7$&$27.22$&$35.94$&$\mathbf{9.18}$&$1.10$&$\mathbf{21.30}$&$6.80$&$0.05$&$0.036$&$0.001$\\
\hline
\hline
\end{tabular}
\end{center}
{\caption[junk]{\label{tab:ch3data}{Submillimetre observations and basic data 
for the sample of eighteen radio quasars from \citet{rawlings} and 
\citet{wsubmm}. The values for the $\rm{[OIII]}$ line strength have
been calculated by various methods (see Sec.~\ref{sec:data}).
Submillimetre detections at the $\ge 2\sigma$ level are given in bold type. 
SC is the assumed fractional synchrotron contamination at 850 $\mu$m. $D$
is the projected linear size of the radio source and a $\dagger$ denotes that is has been measured
between the core and an extended component on just one side of the core. 
The synchrotron contaminations, expressed as fractions of observed $S_{850}$, were taken from
\citet{rawlings}. The parameters $\alpha$ and $\beta$ have been derived using the method of
GRW.
}}}
\end{table*}

Table~\ref{tab:ch3data} shows the basic data for the sample, 
including the $\alpha$ and $\beta$ values derived following the 
prescription of GRW. Submillimetre luminosity is used as a measure of the
radiation from dust, which may be heated by star-formation and/or the central quasar nucleus.
However, since the 6CE and 3CR samples include high-flux-density radio sources, 
it is necessary to attempt to correct for contamination of the submillimetre radiation
by the synchrotron radiation responsible for radio emission.
The fractional corrections for synchrotron 
contamination, included in Table~\ref{tab:ch3data}
for the 6CE and 3CR
sub-samples, were taken from \citet{rawlings} and \citet{wsubmm}. Throughout this 
paper, submillimetre flux densities are converted to $L_{850}$ by, first, correcting 
for synchrotron contamination and then, second, by
assuming a grey body emission spectrum with a dust 
temperature $T_{\rm d} = 40 ~ \rm K$ and emissivity index $\beta_{\rm d} \approx 2$ \citep{pridmc01}. 

\begin{figure*}
\begin{center}
\setlength{\unitlength}{1mm}
\begin{picture}(150,150)
\put(-20,-20){\includegraphics{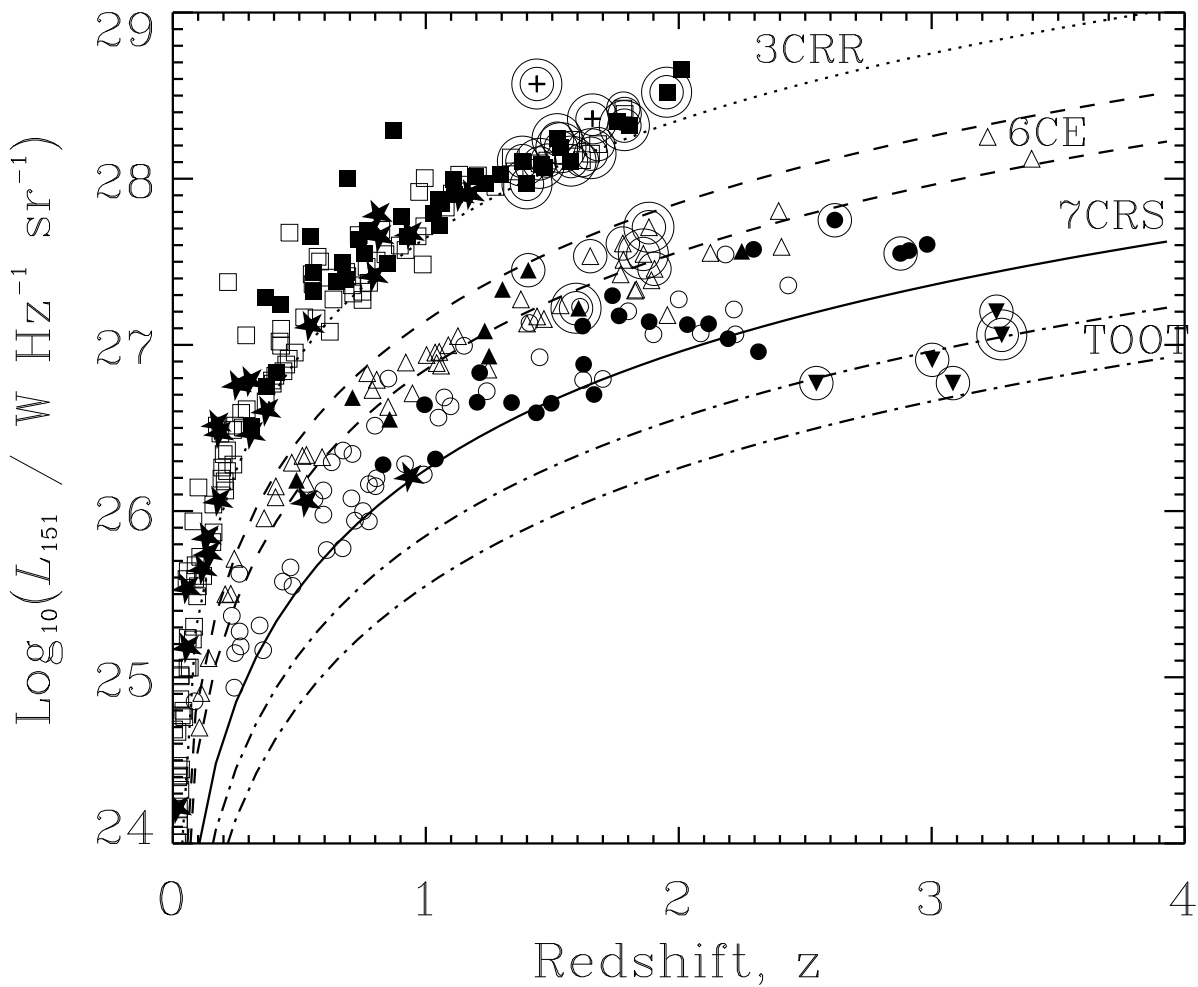}}
\end{picture}
\end{center}
{\caption[junk]{\label{fig:pzch3}
{
The $151$ MHz luminosity $L_{151}$ versus redshift $z$ plane for the 3CRR, 6CE and 
7CRS complete samples with the symbols identifying radio quasars and
radio galaxies from different samples: 3CRR quasars (filled squares);
3CRR radio galaxies (open squares); 6CE quasars (filled triangles); 
6CE radio galaxies (open triangles); 7CRS quasars (filled circles); and 7CRS radio 
galaxies (open circles). Weak quasars (see GRW) are identified by star symbols. 
Also shown are the five TOOT quasars (inverted triangles), and the two 3CR quasars 
(plus signs) described in Sec.~\ref{sec:data}. Circles surrounding symbols show 
objects observed by SCUBA, and additional larger circles 
show objects detected by SCUBA at the $> 2 \sigma$ level.
The dotted line shows the flux-density lower limit for the 3CRR sample, 
the dashed lines show the upper and lower limits for 6CE, 
the solid line shows the lower limit for 7CRS and the dot-dashed lines 
show the upper and lower limits for the TOOT sample (all assuming a radio spectral
index of $0.8$).
}}}
\end{figure*}

Fig.~\ref{fig:pzch3} shows the objects in the sample as a function of $L_{151}$ and 
$z$, in comparison to the full 3CRR/6CE/7CRS dataset used in
GRW. In this paper we are clearly focusing on smaller samples at
higher redshifts, 
but we still span a wide range of $L_{151}$ because of the inclusion of TOOT data.
 
\section{The $L_{850} - \alpha$ correlation}
\label{sec:ch3Lacor}

GRW encoded the $L_{\rm [OIII]} - L_{151}$ correlation for
radio sources using the parameters $\alpha$ and $\beta$, derived from
a principal components analysis. In this section the
correlations between submillimetre ($850$-$\mu$m) luminosity $L_{850}$, 
$L_{151}, L_{\rm [OIII]}$ and $\alpha$ are investigated. The aim
is to find the best way of incorporating submillimetre data into the GLFs derived 
by GRW.

\subsection{New PCA Analysis}
\label{sec:PCA}
A new principal components analysis (PCA) 
was carried out to determine the nature of the 
correlation
between $L_{151}$, $L_{\rm [OIII]}$ and $L_{850}$. 
This analysis used the data of Sec.~\ref{sec:data}.
Although the overall fraction of submillimetre detections is low amongst the fainter
radio samples, such objects have been detected in a statistical sense
\citep{rawlings}. Therefore, we replaced any synchrotron-corrected values 
of $S_{850}$ below the $1\sigma$ detection limit by this limit for the purposes of calculating
$L_{850}$. The
mean values of $\log_{10}L_{151}, \log_{10}L_{\rm [OIII]}$ and $\log_{10}L_{850}$ 
are $27.73, 36.65$ and 
$22.68$ respectively, with associated standard deviations $0.60, 0.55$ and $0.38$.
The results of the PCA are given in Table \ref{tab:ch3pca}. The dominant correlation
(accounting for $\simeq \frac{2}{3}$ of the scatter) is a positive correlation 
between $L_{850}$, $L_{\rm [OIII]}$ and $L_{151}$.
To test if this $1\sigma$-upper-limit approach yielded serious problems, 
we repeated the analysis with ten-times lower values of $\log_{10}L_{850}$ for
the non-detections, which yielded only small changes in the derived
values. Note, however, that this would increase the dispersion in $\log_{10}L_{850}$, so that 
the interpretation of the normalized quantities would be different.

\begin{table}
\begin{center}
\begin{tabular}{lrrr}
\hline\hline
&\multicolumn{1}{c}{$\lambda_{\alpha'}$}&\multicolumn{1}{c}{$\lambda_{\beta'}$}&\multicolumn{1}{c}{$\lambda_{\gamma'}$}\\
\hline
Eigenvalue&$1.93$&$0.57$&$0.50$ \\
Proportion&$0.64$&$0.19$&$0.17$ \\
\hline
&\multicolumn{1}{c}{$\mathbf{e_{\alpha'}}$}&\multicolumn{1}{c}{$\mathbf{e_{\beta'}}$}&\multicolumn{1}{c}{$\mathbf{e_{\gamma'}}$}\\
\hline
$\log_{10}(L_{151})_{\rm{norm}}$&$0.593$&$ -0.024$&$0.805$\\
$\log_{10}(L_{\rm [OIII]})_{\rm norm}$&$0.571$&$-0.693$&$-0.441$\\
$\log_{10}(L_{850})_{\rm{norm}}$&$0.569$&$0.720$&$-0.397$ \\
\hline\hline
\end{tabular}
\caption[PCA table]{\label{tab:ch3pca}{Table of the eigenvalues $\lambda_{\alpha'}$, 
$\lambda_{\beta'}$ and $\lambda_{\gamma'}$, and 
eigenvectors $\mathbf{e_{\alpha'}}$, $\mathbf{e_{\beta'}}$ and $\mathbf{e_{\gamma'}}$
from the PCA analysis, such that 
$\mathbf{e_{\alpha'}}$ = $0.593 \, \log_{10}(L_{151})_{\rm norm} 
+ 0.571 \, \log_{10}(L_{\rm [OIII]})_{\rm norm} +
 0.569 \, \log_{10}(L_{850})_{\rm{norm}}$ etc. The components are normalized in the
same manner as GRW and primes are 
used to distinguish these principal components from those found in their 
previous, two-variable, PCA analysis.
}}
\end{center}
\end{table}

\subsection{Survival analysis}
\label{sec:asurv}
The new PCA (Sec.~\ref{sec:PCA})
has yielded the basic result that $L_{850}$ is positively correlated with 
$L_{151}$ and $L_{\rm [OIII]}$. As the $\alpha$ parameter from GRW
encodes the $L_{151}$ and $L_{\rm [OIII]}$
correlation, it is expected that $L_{850}$ will be correlated with $\alpha$. 
In this section, a survival analysis is used to quantify the correlation between 
$\log_{10}L_{850}$ and the 
$\log_{10}L_{151}, \log_{10}L_{\rm [OIII]}$ combination through $\alpha$, now taking   
into account the $2\sigma$ upper limits which are important because nearly half of the objects
are not detected at the $2\sigma$ level at $850 \mu$m.
A correlation and regression analysis was carried out using the {\sc Asurv} software
\citep{ifn86}.

\begin{figure*}
\begin{center}
\setlength{\unitlength}{1mm}
\begin{picture}(150,60)
\put(-15,-5){\includegraphics{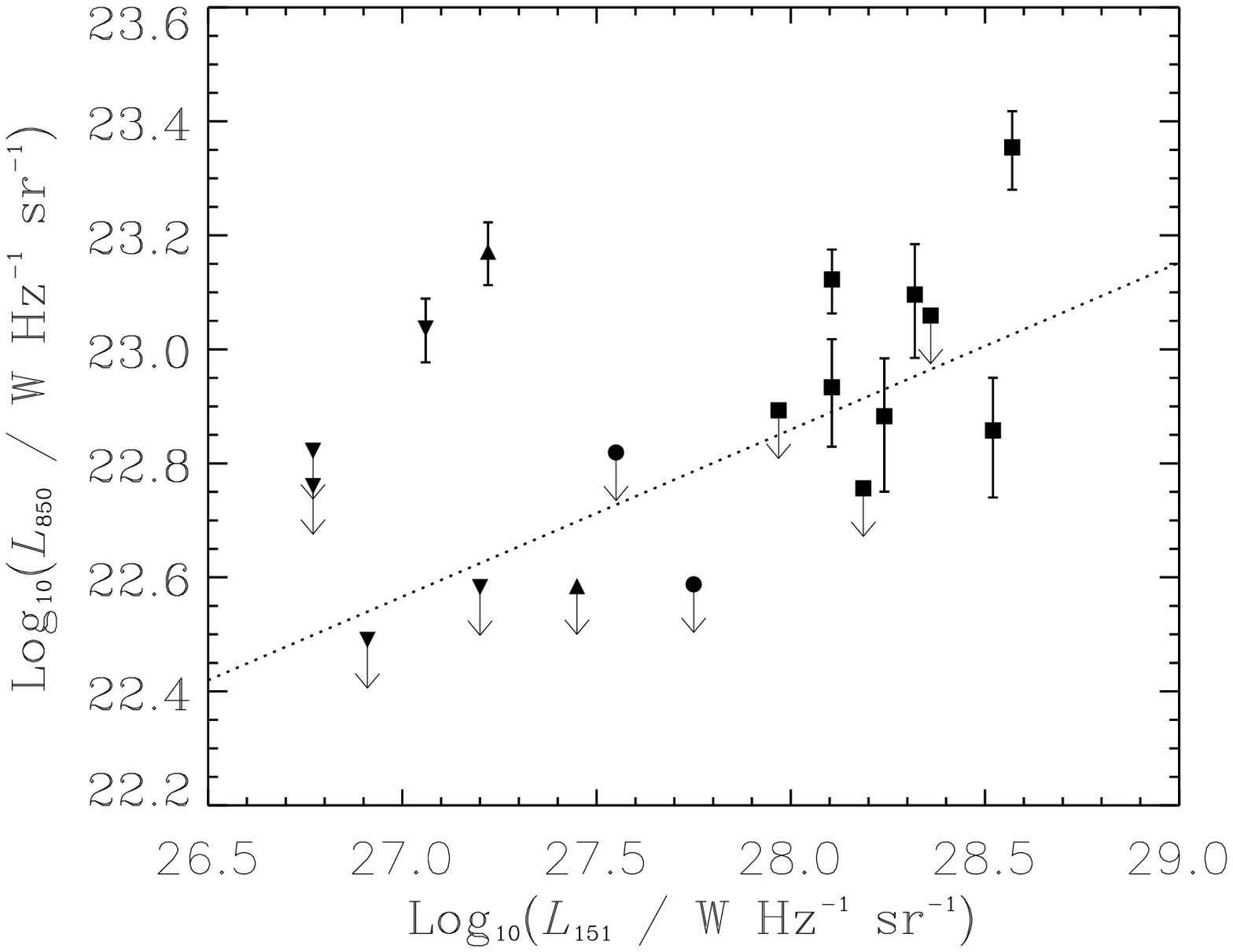}}
\put(45,-5){\includegraphics{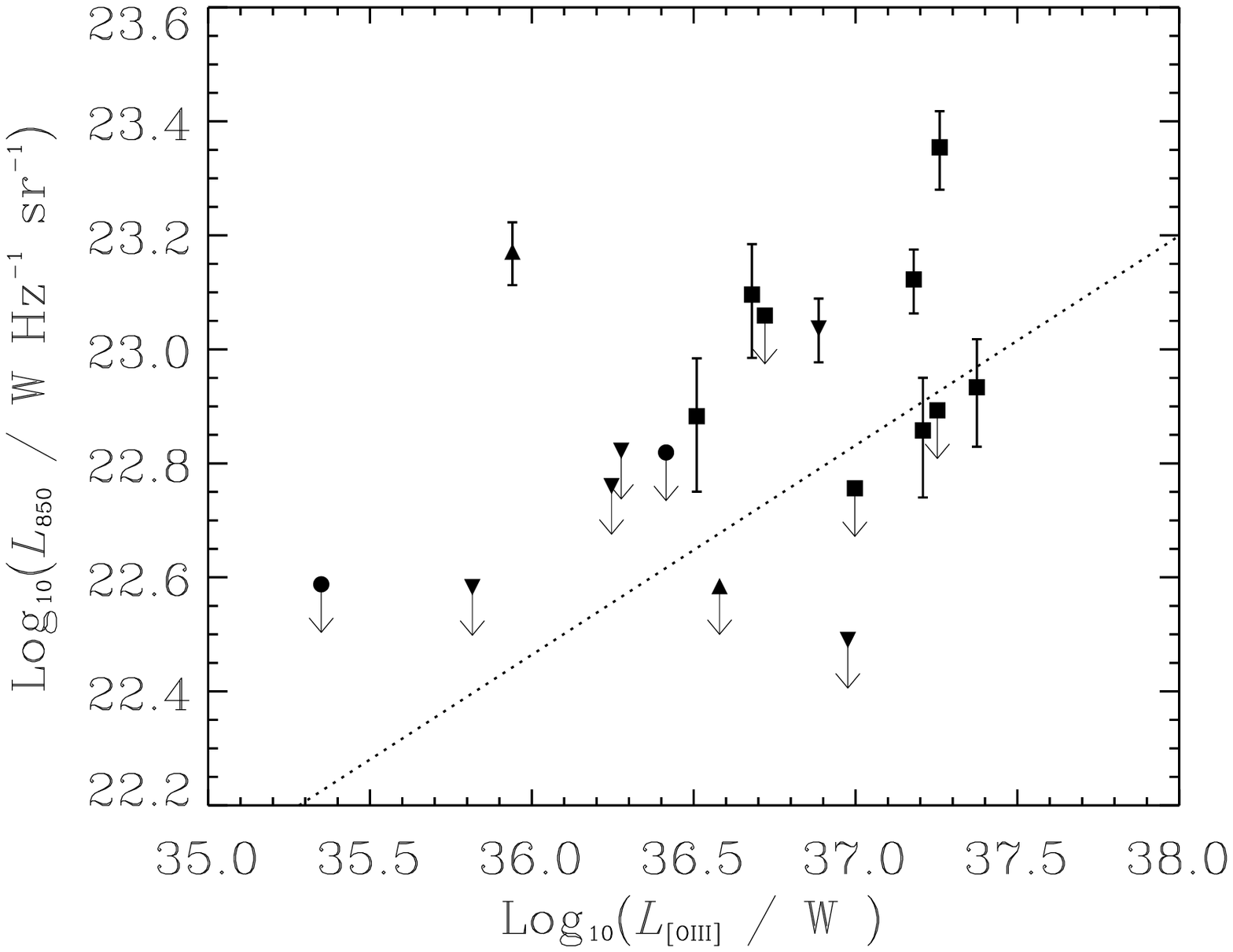}}
\put(105,-5){\includegraphics{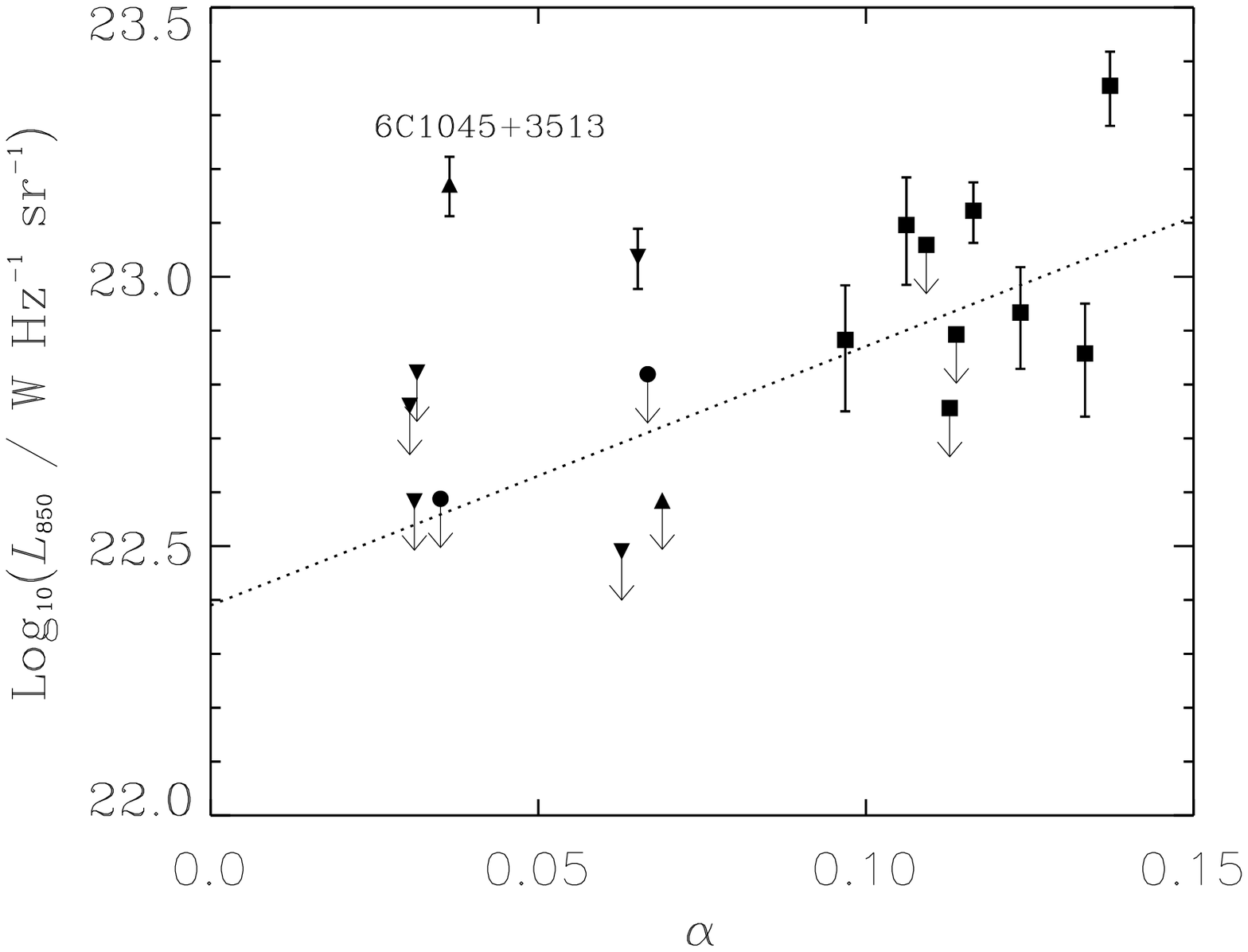}}
\end{picture}
\end{center}
\vspace{-0.5cm}
{\caption[junk]{\label{fig:LLcor}
{
The correlations of $\log_{10}L_{850}$ with $\log_{10}L_{151}$,
$\log_{10}L_{\rm [OIII]}$ and $\alpha$ for the sample 
of eighteen quasars defined in Table~\ref{tab:ch3data}. The 
$2\sigma$ upper limits are shown for the non-detections
(symbols as in Fig.~\ref{fig:pzch3}).
The best-fitting (dotted) lines are shown from an 
analysis using the Buckley--James method.
}}}
\end{figure*}

\begin{table*}
\begin{center}
\begin{tabular}{l|cc|ccc|cccc}
\hline\hline
Correlation of&\multicolumn{2}{c}{Cox Prop Hazard}&\multicolumn{3}{c}{BHK test}&\multicolumn{4}{c}{Buckley--James Method}\\
$\log_{10}L_{850}$ with&$\chi^2$&$\rm{P_c}$&$\tau$&Z&$\rm{P_k}$&Intercept&Gradient&SD&SDR\\
\hline
$\log_{10}L_{151}$&$3.632$&$0.057$&$0.523$&$1.744$&$0.081$&$14.65$&$0.29$&$0.17$&$0.26$\\
$\log_{10}L_{\rm [OIII]}$&$2.459$&$0.117$&$0.444$&$1.482$&$0.139$&$9.23$&$0.21$&$0.21$&$0.27$\\
$\alpha$&$4.352$&$0.037$&$0.575$&$1.917$&$0.055$&$22.39$&$4.81$&$2.86$&$0.27$\\
\hline\hline
\end{tabular}
{\caption[junk]{\label{tab:correl}{
The correlation of $\log_{10}L_{850}$ with $\log_{10}L_{151}$, 
$\log_{10}L_{\rm [OIII]}$ and $\alpha$ using Cox's Proportional Hazard model, the 
generalized Kendall's rank correlation test (BHK test) and Buckley--James method for 
regression
with the {\sc Asurv} program \citep{ifn86}. 
$\chi^2$ is the global $\chi^2$, and $\rm{P_c}$ is the probability that a correlation
is not present for Cox's Proportional Hazard model. $\tau$ is Kendall's $\tau$, Z is the
Z-value and $\rm{P_K}$ is the probability that a correlation
is not present for the BHK test.
SD is the standard deviation of the
gradient and SDR is the standard deviation of regression for the Buckley--James method. 
}}}
\end{center}
\end{table*}

Table \ref{tab:correl} shows the correlation coefficients using three methods:
Cox's Proportional Hazard model, the generalized Kendall's rank correlation test for 
censored data sets using the BHK test, both of 
which measure the
degree of association of two variables, and the Buckley--James method which is the usual
least squares regression adapted for censored data.
Fig.~\ref{fig:LLcor} shows the correlations of $L_{850}$ with
$L_{151}$ and $L_{\rm [OIII]}$, and also the correlation of 
$L_{850}$ with $\alpha$. 

This battery of statistical tests suggets that there are real, positive 
correlations between the variables considered. However, the 
best-fit lines are strongly dependent on how the upper limits are
treated. In the next section we will extend the GLFs from GRW to encode the
$L_{850}-\alpha$ correlation and to treat upper limits in a way more appropriate to the 
problem under consideration.

\section{Generalized luminosity functions}
\label{sec:ch3glf}
\subsection{Definition}

The logical extension to the GLF $\rho_{\rm Q}(L_{151}, L_{\rm [OIII]}, z) =  
\rho_{\rm Q}(\alpha, \beta, z)|J|$, where $J$ is the appropriate Jacobian defined
by GRW,
is to incorporate submillimetre data and define a GLF $\phi$ which predicts
the number of radio quasars per unit comoving 
volume (in units of Mpc$^3$) per unit base-ten logarithm
of $L_{151}$, $L_{\rm [OIII]}$ and $L_{850}$ at a given redshift $z$. 
The PCA (Sec.~\ref{sec:PCA}) and survival analyses (Sec.~\ref{sec:asurv}) 
have demonstrated that the $L_{850}$--$\alpha$ correlation can be encoded as a linear relation 
\mbox{$\log_{10}L_{850}(\alpha)  = G \alpha + C$}, where $G$ is the gradient and 
$C$ is the intercept, plus scatter about this relation. We encode the scatter using a
Gaussian distribution function about $L_{850}(\alpha)$, with a free parameter 
$\sigma_L$, so that

\begin{eqnarray}
\label{eqn:ch3glf}
\nonumber
\lefteqn{\phi(L_{151}, L_{\rm [OIII]}, L_{850}, z)} \\
\nonumber
&=& \frac{{\rm d}^{4} N}{{\rm d}\log_{10}L_{151}{\rm d}\log_{10}L_{\rm [OIII] }{\rm d}\log_{10}L_{850}{\rm d}V} \\  
&=& \rho(L_{151}, L_{\rm [OIII], z})  f(L_{850}, \alpha, \sigma_L),
\end{eqnarray}
\noindent where 

\begin{eqnarray}
\nonumber
\lefteqn{f(L_{850}, \alpha, \sigma_L)} \\
&=&\frac{1}{\sigma_L \sqrt{2 \pi}} \exp{\left( - \frac{ [\log_{10}L_{850} 
- \log_{10}L_{850}(\alpha)]^2}{2\sigma_L^2} \right)},
\end{eqnarray}
and ${\rm d V}$ represents the comoving volume element at an epoch corresponding to redshift
$z$.

Given the large proportion of data with only upper limits at $850 ~ \mu$m, we need
to adapt this GLF so that it takes non-detections properly
into account. We will assume that an observer will measure a flux density $S_{\rm obs}$
at $850 ~ \mu$m (after correction for synchrotron contamination)
which is drawn from a Gaussian distribution
about the actual flux density $S_{850}$ with the standard deviation given by the 
observational error $\sigma_{\rm obs}$.
Thus, an observed value of flux density $S_{\rm obs}$ could be due to a range of 
real (synchrotron-corrected) flux density $S_{850}$, each of which corresponds to a luminosity $L_{850}$ 
\citep[assuming $\beta_{\rm d}$ and $T_d$ from][]{pridmc01}.
In this way, we define a new GLF $\Phi$ in terms of $S_{\rm obs}$ and 
observational error $\sigma_{\rm obs}$, related to the GLF $\phi$ by 

\begin{eqnarray}
\lefteqn{\Phi(L_{151}, L_{\rm [OIII]}, S_{\rm obs}, \sigma_{\rm{obs}}, z)} \\
&=& \int \phi(L_{151}, L_{\rm [OIII]}, L_{850}, z) g(S_{\rm{obs}}, S_{850}, 
\sigma_{\rm{obs}}) |J'| {\rm d} S_{850},
\nonumber
\end{eqnarray}

\noindent where

\begin{eqnarray}
g(S_{\rm{obs}}, S_{850}, \sigma_{\rm{obs}}) &=& {\rm Prob}(S_{\rm obs} | S_{850}, \sigma_{\rm obs})\\
&=& \frac{1}{\sigma_{\rm obs}\sqrt{2 \pi} } 
\exp{\left(- \frac{[S_{\rm{obs}} - S_{850}]^2}{2 \sigma_{\rm obs}^2}\right)},
\nonumber
\end{eqnarray}

\noindent
and $|J'|$ is the new Jacobian needed to convert from $S_{850}$ to $\log_{10}L_{850}$.

\subsection{Normalization corrections}
\label{sec:skyarea}
One of the known small problems with the GLFs defined by GRW
is that they predict too few objects at low redshift, and correspondingly too many objects 
at high redshifts in comparison to the 3CRR, 6CE and 7CRS data. This was a 
consequence of trying to fit the 6C and 7C source counts (in addition to the 3CRR, 6CE 
and 7CRS redshift surveys) and may reflect imperfections in the 
evolutionary model used for the high-$\alpha$ population (GRW).
Since the redshift range of our sample 
is $1.37 < z < 3.5$, and we have only a small sample of objects, we expect that such
imperfections may be important. Table \ref{tab:fudgefactors} shows how 
we have attempted to correct for this effect by scaling the sky areas of each 
subsample by the ratio of the total number of objects in each subsample
to the nearest whole number of objects (quasars and radio galaxies) predicted by 
the unified schemes for each of the models (2RS, 2SS and 2RR; see Table 4 of GRW). 
Note that we do not yet 
know the number of radio galaxies in the TOOT surveys in this redshift range, so 
we have estimated the correction
factor by using the average quasar fraction, from the three models, 
to predict the number of TOOT radio galaxies from the number of TOOT quasars.

\begin{table*}
\label{tab:numbers}
\begin{center}
\begin{tabular}{lrrrrrrrrrr}
\hline\hline
&\multicolumn{2}{c}{3CRR}&\multicolumn{2}{c}{6CE}&\multicolumn{2}{c}{7CRS}&\multicolumn{2}{c}{TOOT}&\multicolumn{2}{c}{Total}\\
&\multicolumn{1}{c}{Q}&\multicolumn{1}{c}{RG}&\multicolumn{1}{c}{Q}&\multicolumn{1}{c}{RG}&\multicolumn{1}{c}{Q}&\multicolumn{1}{c}{RG}&\multicolumn{1}{c}{Q}&\multicolumn{1}{c}{RG}&\multicolumn{1}{c}{Q}&\multicolumn{1}{c}{RG}\\
\hline
\hline
Data&\multicolumn{1}{c}{$9$}&\multicolumn{1}{c}{$7$}&\multicolumn{1}{c}{$2$}&\multicolumn{1}{c}{$5$}&\multicolumn{1}{c}{$2$}&\multicolumn{1}{c}{$0$}&\multicolumn{1}{c}{$5$}&\multicolumn{1}{c}{?}&\multicolumn{1}{c}{$18$}&\multicolumn{1}{c}{?}\\
\hline
2RS&$12.38$&$8.23$&$7.13$&$5.18$&$2.64$&$1.92$&$7.88$&$21.24$&$32.05$&$38.04$\\
2SS&$8.24$&$12.23$&$7.71$&$5.39$&$2.58$&$1.87$&$11.37$&$17.75$&$31.06$&$39.27$\\
2RR&$12.65$&$7.97$&$7.41$&$5.37$&$2.58$&$1.88$&$8.26$&$ 20.86$&$33.12$&$37.19$\\
\hline
2RS&$9.43$&$6.27$&$4.53$&$2.31$&$0.78$&$1.28$&$4.35$&$11.72$&$19.07$&$21.99$\\
2SS&$6.28$&$9.32$&$4.72$&$2.26$&$0.81$&$1.25$&$6.27$&$9.79$&$18.37$&$22.85$\\
2RR&$9.64$&$6.07$&$4.70$&$2.26$&$0.81$&$1.25$&$4.56$&$11.51$&$19.70$&$21.49$\\
\hline
\hline
\end{tabular}
{\caption[junk]{\label{tab:fudgefactors}{
Table of the number of radio quasars (Q) and radio galaxies (RG) from each survey 
in the sample (upper panel) and the predicted numbers of quasars and radio galaxies 
in each subsample before and after (middle and lower panels respectively) a factor was 
applied to the sky areas to correct for the imperfections in 
the GLFs at high redshifts (Sec.~\ref{sec:skyarea}); 
see GRW for a full description of the three models (2RS, 2SS, 2RR) which
represent the three different unified schemes. Note that the number of radio galaxies in the
relevant redshift range is not yet established for the TOOT survey.
}}}
\end{center}
\end{table*}

\subsection{Results}
\label{sec:ch3results}

The GLF of Eqn.~\ref{eqn:ch3glf} was combined with each of the 
three models (2RS, 2SS and 2RR) defined in Table 4 of GRW using
the best-fitting unified scheme parameters found in their
Table 5.
A maximum-likelihood analysis was performed in order to find best-fitting values of
$\sigma_L$, $G$ and $C$. The likelihood function $\eta$ was defined as

\begin{eqnarray}
\label{eqn:Sch3}
\nonumber
\eta &=& - 2 \sum_{i=0}^{18} \ln [
\Phi(L_{151_i}, L_{\rm [OIII]_i}, S_{\rm obs_i}, \sigma_{\rm{obs_i}}, z_i)] \\
\nonumber
&+& 2 \int \!\!\! \int \!\!\! \int \!\!\! \int  [\phi(L_{151}, L_{\rm [OIII]}, L_{850}, z) \, \Omega(L_{151}, z) \\
&\times& \frac{{\rm d}V}{{\rm d}z} \, {\rm d}z \, {\rm d} (\log_{10}L_{151}) \, {\rm d} 
(\log_{10}L_{\rm [OIII]}) \, {\rm d} (\log_{10}L_{850}) ], 
\end{eqnarray}

\noindent 
where $\Omega$ is the sky area available for each sample of radio sources, 
modified by the factors derived from Table~\ref{tab:fudgefactors}.

The best-fitting values of the three free parameters $\sigma_L$, $G$ and $C$ were found
to be almost identical for each of the three models and are presented in 
Table~\ref{tab:ch3mainres}. 
The steeper gradient and different normalization of this line compared to that found by
the survival analysis can be explained by our new method of treating upper limits
which means that non detections can be drawn from a wide range of underlying 
submillimetre luminosities.

\begin{figure}
\begin{center}
\setlength{\unitlength}{1mm}
\begin{picture}(150,80)
\put(0,0){\includegraphics{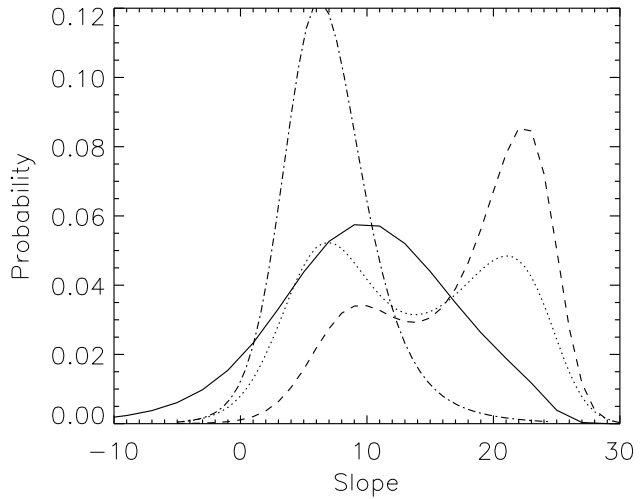}}
\end{picture}
\end{center}
\vspace{-1.5cm}
{\caption[junk]{\label{fig:Pslope2}{Marginalized probability distributions for the gradient $G$ of the
$\log_{10}L_{850} - \alpha$ correlation derived from the 2RS GLF (similar probability
distributions can be inferred using models 2RR and 2SS). Different datasets have been
used: the radio quasars from Table~\ref{tab:ch3data} (dot-dashed line); 
dataset A (dotted line); B (solid line); and C (dashed line). Datasets A, B and C are defined in
Sec.~\ref{sec:ch3compwill}.
.}}}
\end{figure}

A probability distribution was found for the gradient $G$ of the $L_{850}$ versus 
$\alpha$ correlation by marginalising 
the likelihood (the function $\eta$ in Eqn.~\ref{eqn:Sch3})
over $C$ and $\sigma_L$, and normalizing. Fig.~\ref{fig:Pslope2} shows that
the probability that there is a positive correlation ($S > 0$) between $L_{850}$
and $\alpha$ is $\sim 99$ per cent. Note that this is a different conclusion than
one would draw from the `formal' error from Table~\ref{tab:ch3mainres} because
marginalization allows correctly for the complicated shape of the probability 
distributions. Note, however, that there is still quite a large error on the gradient $G$
of the correlation. 

The relative probability of the models with a receding torus in the high-$\alpha$
population (models 2RS and 2RR)
with respect to the GLF with a non-luminosity-dependent unified scheme in both 
populations (model 2SS) is $\sim 100$. The main reason for this is that the adoption of a
receding torus means that the value of the transition angle $\Theta_{\rm trans}$ increases
with $\alpha$, and as all the radio quasars in the sample have high values of $\alpha$ 
(Fig.~\ref{fig:LLcor}), Eqn.~10 of GRW implies a corresponding boost in the 
values of $\rho_{\rm Q}$ [note that 
the free parameter fixing the behaviour of $\Theta_{\rm trans}$ has not been allowed
to vary in our new likelihood analysis]. However, in complete samples,
these increased likelihoods of seeing radio quasars goes
hand-in-hand with decreased likelihoods of seeing radio galaxies
(see Fig.\ 8 of GRW), so we defer interpretation of any difference in likelihoods until, in 
Sec.~\ref{sec:ch3compwill}, we consider complete samples including both 
radio quasars and radio galaxies.

\begin{table*}
\label{tab:parameters}
\begin{center}
\begin{tabular}{lcccccc}
\hline\hline
\multicolumn{1}{c}{Model}&$\sigma_L$&$G$&$C$&$\eta_{\rm min}$&$\det(\nabla \nabla \eta_{\rm min})$&$P_{\rm 2RS}$\\
\hline
2RS&$0.365_{-0.101}^{+0.138}$&$6.036 \pm 5.180$&$22.136 \pm 0.493$&$651.79$&$1534.28$&$1.000$\\
2SS&$0.365_{-0.101}^{+0.138}$&$6.036 \pm 5.180$&$22.136 \pm 0.493$&$661.37$&$1534.27$&$0.009$\\
2RR&$0.365_{-0.101}^{+0.138}$&$6.036 \pm 5.180$&$22.136 \pm 0.493$&$652.04$&$1534.28$&$0.886$\\
\hline\hline
\end{tabular}
\end{center}
\caption{\label{tab:ch3mainres} 
Results of the likelihood analysis of Sec.~\ref{sec:ch3results} which is based on
the sample of radio quasars in Table~\ref{tab:ch3data}.
The best-fitting values of the parameters $\sigma_L$, $G$ 
and $C$, the minimum value of the likelihood function $\eta_{\rm min}$ corresponding
to these best-fitting values and $\det(\nabla \nabla \eta_{\rm min})$ for 
the three unified schemes. `Formal' error bars on the parameters are 
obtained from the $\nabla \nabla \eta$ matrix in the normal manner (see GRW), and 
the probability $P_{\rm 2RS}$ of each of the models with
respect to model 2RS is calculated using Equation 12 of GRW.}
\end{table*}

\section{Comparison with previous studies}
\label{sec:ch3comp}

\subsection{Comparison with the Willott {\it et al.} sample}
\label{sec:ch3compwill}

\citet{wsubmm} found that radio quasars were a 
factor of $\sim 5$ times brighter than
radio galaxies in the submillimetre, based on SCUBA observations of eleven 3CRR, 3CR and 6CE
quasars (see the lowest rows of Table~\ref{tab:ch3data})
and twelve 3CRR and 6CE radio galaxies (Table~\ref{tab:rg}).
This factor could be reduced somewhat by adopting larger estimates of
synchrotron contamination, but not below a factor $\sim$2, and so not to the
point where radio quasars and radio galaxies have comparable dust
luminosities. We next make a comparison of the ability of the three GLFs to reproduce
the distribution of submillimetre fluxes of these data. 
A one-dimensional optimization algorithm \citep[the Golden Section Search method; ][]{press} 
was used to find the best-fitting value of $\sigma_L$,
using the data (on both radio quasars and radio galaxies) presented by \citet{wsubmm}.
As these radio quasars and radio galaxies now span a much smaller range in $\alpha$,
compared to the full sample used to constrain the gradient $G$ and offset $C$ 
(Sec.~\ref{sec:ch3results}), we initially kept the best-fitting values of $G$ and $C$ constant 
during this optimization.

\begin{table*}
\begin{center}
\small
\begin{tabular}{lccrccr@{ $\pm$ }rr@{ $\pm$ }rcrr}
\hline\hline
\multicolumn{1}{l}{ Name}&\multicolumn{1}{c}{Class}&\multicolumn{1}{c}{Redshift}
&\multicolumn{1}{c}{$D /$}&\multicolumn{1}{c}{$\log_{10}[L_{151} / $}&\multicolumn{1}{
c}{$\log_{10}[L_{\rm [OIII]} /$}&\multicolumn{2}{c}{$S_{850} /$}&\multicolumn{2}{c}{$
S_{450} /$}&\multicolumn{1}{c}{SC}&\multicolumn{1}{c}{$\alpha$}&
\multicolumn{1}{c}{$\beta$}\\
&&$z$&\multicolumn{1}{c}{kpc}&\multicolumn{1}{c}{$\rm W Hz^{-1} sr^{-1}]$}&
\multicolumn{1}{c}{W}]&\multicolumn{2}{c}{mJy}&\multicolumn{2}{c}{mJy}&\multicolumn{1}{
c}{}&&\\
\hline\hline
4C~13.66&G&$1.450$&$50.7$&$28.09$&$36.63$&$\mathbf{3.53}$&$0.96$&$-16.20$&$18.20$
&$0.00$&$0.095$&$0.010$\\
3C~329&G&$1.781$&$94.6$&$28.42$&$36.38$&$0.83$&$1.00$&$-2.10$&$18.30$&$0.00$&$0.0
99$&$0.031$\\
3C~241&G&$1.617$&$7.7$&$28.11$&$36.75$&$1.81$&$0.94$&$14.80$&$12.60$&$0.00$&$0.10
1$&$0.006$\\
3C~294&G&$1.786$&$126.7$&$28.39$&$36.97$&$0.19$&$0.78$&$5.40$&$13.40$&$0.00$&$0.1
20$&$0.008$\\
3C~322&G&$1.681$&$279.4$&$28.21$&$35.97$&$-0.05$&$1.06$&$-37.50$&$16.00$&$0.00$&$0
.076$&$0.038$\\
3C~437&G&$1.480$&$291.0$&$28.11$&$36.37$&$-1.18$&$0.98$&$2.90$&$17.30$&$0.00$&$0.0
86$&$0.020$\\
3C~470&G&$1.653$&$203.2$&$28.15$&$35.93$&$\mathbf{5.64}$&$1.08$&$57.80$&$32.90$&$0
.00$&$0.072$&$0.037$\\
6C~0820+3642&G&$1.860$&$193.8$&$27.55$&$36.01$&$\mathbf{2.07}$&$0.96$&$13.60$&$18.
00$&$0.00$&$0.052$&$0.011$\\
6C~0901+3551&G&$1.904$&$21.9$&$27.46$&$36.13$&$-1.83$&$1.15$&$-19.30$&$8.20$&$0.00
$&$0.053$&$0.003$\\
6C~0905+3955&G&$1.882$&$934.5$&$27.71$&$35.77$&$\mathbf{3.62}$&$0.89$&$31.20$&$16.
20$&$0.00$&$0.049$&$0.026$\\
6C~0919+3806&G&$1.650$&$88.1$&$27.53$&$35.93$&$-0.88$&$1.05$&$10.50$&$10.10$&$0.00
$&$0.048$&$0.014$\\
6C~1204+3708&G&$1.779$&$435.2$&$27.61$&$36.45$&$0.16$&$1.25$&$45.10$&$26.60$&$0.00
$&$0.070$&$-0.003$\\
\hline
\hline
\end{tabular}
\end{center}
{\caption[junk]{\label{tab:rg}{Submillimetre observations and basic data
for the sample of twelve radio galaxies from \citet{wsubmm}.
The values for the $\rm{[OIII]}$ line strength have
been calculated by various methods (see Sec.~\ref{sec:data}).
Submillimetre detections at the $\ge 2\sigma$ level are given in bold type. 
SC is the assumed fractional synchrotron contamination at 850 $\mu$m. 
$D$
is the projected linear size and a $\dagger$ denotes that it is measured between
the core and an extended component on just one side of the core.
contaminations, expressed as fractions of observed $S_{850}$, were taken from
\citet{rawlings}; $\alpha$ and $\beta$ are the parameters defined by GRW.
}}}
\end{table*}

\begin{figure}
\begin{center}
\setlength{\unitlength}{1mm}
\begin{picture}(150,80)
\put(0,0){\includegraphics{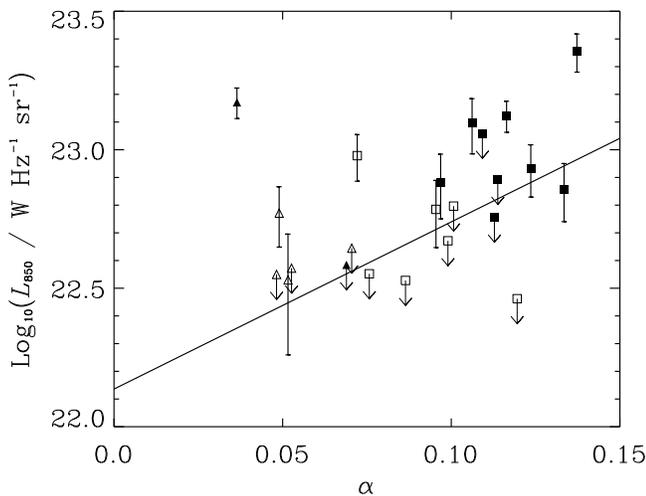}}
\end{picture}
\end{center}
\vspace{-1.5cm}
{\caption[junk]{\label{fig:al850rg}
{The $850 \mu$m luminosity $L_{850}$ against $\alpha$ for the sample 
of eleven quasars (last 11 rows of Table~\ref{tab:ch3data}) and twelve radio galaxies 
(Table~\ref{tab:rg}) considered 
by \cite{wsubmm} with 
symbols as in Fig.~\ref{fig:pzch3}. 
$2\sigma$ upper limits are shown for the non-detections.
The best-fitting line from 
the optimization procedure is shown (solid line; see Sec.\ref{sec:ch3compwill}).
}}}
\end{figure}

\begin{table}
\begin{center}
\begin{tabular}{lcccc}
\hline\hline
\multicolumn{1}{c}{Model}&$\sigma_L$&$\eta_{\rm min}$&$\det(\nabla\nabla \eta_{\rm min})$&$P_{\rm 2RS}$\\
\hline
2RS&$0.481_{-0.140}^{+0.197}$&$876.88$&$45.2$&$1.000$\\
2SS&$0.481_{-0.140}^{+0.197}$&$884.57$&$45.2$&$0.028$\\
2RR&$0.481_{-0.140}^{+0.197}$&$877.45$&$45.2$&$0.752$\\
\hline\hline
\end{tabular}
\end{center}
\caption{\label{tab:willres} 
Results of the likelihood analysis of Sec.~\ref{sec:ch3compwill} which is based on
the radio quasars and radio galaxies in the sample of \citet{wsubmm}.
The best-fitting parameter $\sigma_L$,
the minimum value of the likelihood function $\eta_{\rm min}$ corresponding 
to this value and $\det(\nabla \nabla \eta_{\rm min})$ for
the three unified schemes. 
`Formal' error bars on the parameters are 
obtained from the $\nabla \nabla \eta$ matrix in the normal manner (see GRW), and
the probability $P_{\rm 2RS}$ of each of the models with
respect to model 2RS is calculated using 
Equation 12 of GRW.}
\end{table}

We present the results in Table \ref{tab:willres}.
Fig.~\ref{fig:al850rg} shows this sample with the best-fitting gradient
determined by the optimization procedure in Sec.~\ref{sec:ch3results}.
Note that the addition of the radio galaxies leads to a small increase in the 
derived scatter $\sigma_L$ in the relationship between $L_{850}$ and $\alpha$, 
but that the scatter is still insufficient to hide what appears to
be a significant difference in the $L_{850}$ values of radio quasars
and radio galaxies.
For this sample, we find that 
models with a receding torus in the high-$\alpha$ population (2RS and 2RR)
are strongly favoured (by a factor of $\sim 40$) over a model with a torus
which has a constant opening angle in the high-$\alpha$ population. 
This constrasts with the very marginal (factor $\sim$ 2)
difference in likelihoods between such models in GRW.
Since, essentially, the $L_{151}$ and $L_{\rm [OIII]}$ data considered are a subset of the
data considered by GRW, and (in contrast to the likelihood analysis of 
Sec.~\ref{sec:ch3results}) both radio quasars and radio galaxies are included,
the reason that we are reaching such a different conclusion is
that this current study focuses on luminosity differences between radio quasars 
and radio galaxies in just the high-$\alpha$ population. The use of 
a relatively narrow
range of redshifts, so that the confusions associated with mixing low- and high-$\alpha$
populations and with wide ranges of redshift is an additional factor, as is
the additional constraint of the submillimetre properties.
\citet{wsubmm} attributed the difference in submillimetre flux-densities 
between radio quasars and radio galaxies to the dust being heated by quasars 
(rather than by starbursts) which would cause emission-line luminosities to be 
correlated with submillimetre luminosities if both reflect the luminosity of
the central source.
This result is consistent with
a receding torus because objects with more luminous emission lines are 
more likely to be viewed as quasars, because of the larger torus opening
angles, and therefore to have higher $L_{\rm [OIII]}$, $\alpha$ 
and $L_{850}$. 

The weighted mean of the synchrotron-corrected submillimetre fluxes for the radio 
quasars in the \citet{wsubmm} sample is $4.72 \pm 0.36$ mJy, and for the radio galaxies it is 
$1.25 \pm 0.29$ mJy; the median is $4.47$ mJy for quasars and $0.51$ mJy for 
radio galaxies. We have simulated distributions of $S_{850}$ from the average of $30$ simulations
using each of the three GLFs. For model 2SS, the shift is $\approx 0$ mJy 
with the median $S_{850}$ being $3$ mJy for both radio quasars and radio galaxies.
This is as expected because, without a receding torus, experiments that 
focus only on the high-$\alpha$ population would find no difference in the $\alpha$ and
hence $S_{850}$ values of radio quasars and radio galaxies.
We find that the shift in $S_{850}$ between the median quasar
and the median radio galaxy is $1.0$ mJy for models 2RR and 2RS 
(the median quasar has a flux density of $3.0$ mJy and the median radio galaxy 
has a flux density of $2.0$ mJy), so that quasars are
predicted to be $1.5$-times brighter in the submillimetre than radio galaxies. 
Encouragingly, then, shifts in the direction seen in the data are predicted, but
not by a large enough factor. It is easy to see how this factor
has arisen. The median $\alpha$ is $0.113$ for quasars and $0.074$ for radio galaxies due a larger
median emission line luminosity of $\log_{10}(L_{\rm [OIII]} / \rm W) = 37.00$ for 
quasars than the corresponding value of $36.25$ for radio galaxies,
with relatively well-matched radio luminosities [median 
$\log_{10}(L_{151} / \rm{(W Hz^{-1} sr^{-1})})$ for quasars of $28.19$ compared to
$28.10$ for radio galaxies]. Using the best-fitting $L_{850}-\alpha$ relation,
this difference in $\alpha$ leads to median $\log_{10}(L_{850} / \rm{W Hz^{-1} sr^{-1}})$ 
of $22.82$ for quasars and $22.58$ for radio galaxies - a difference of $0.24$ or a 
factor of $\approx 1.7$.

\begin{figure*}
\begin{center}
\setlength{\unitlength}{1mm}
\begin{picture}(150,60)
\put(-5,0){\includegraphics{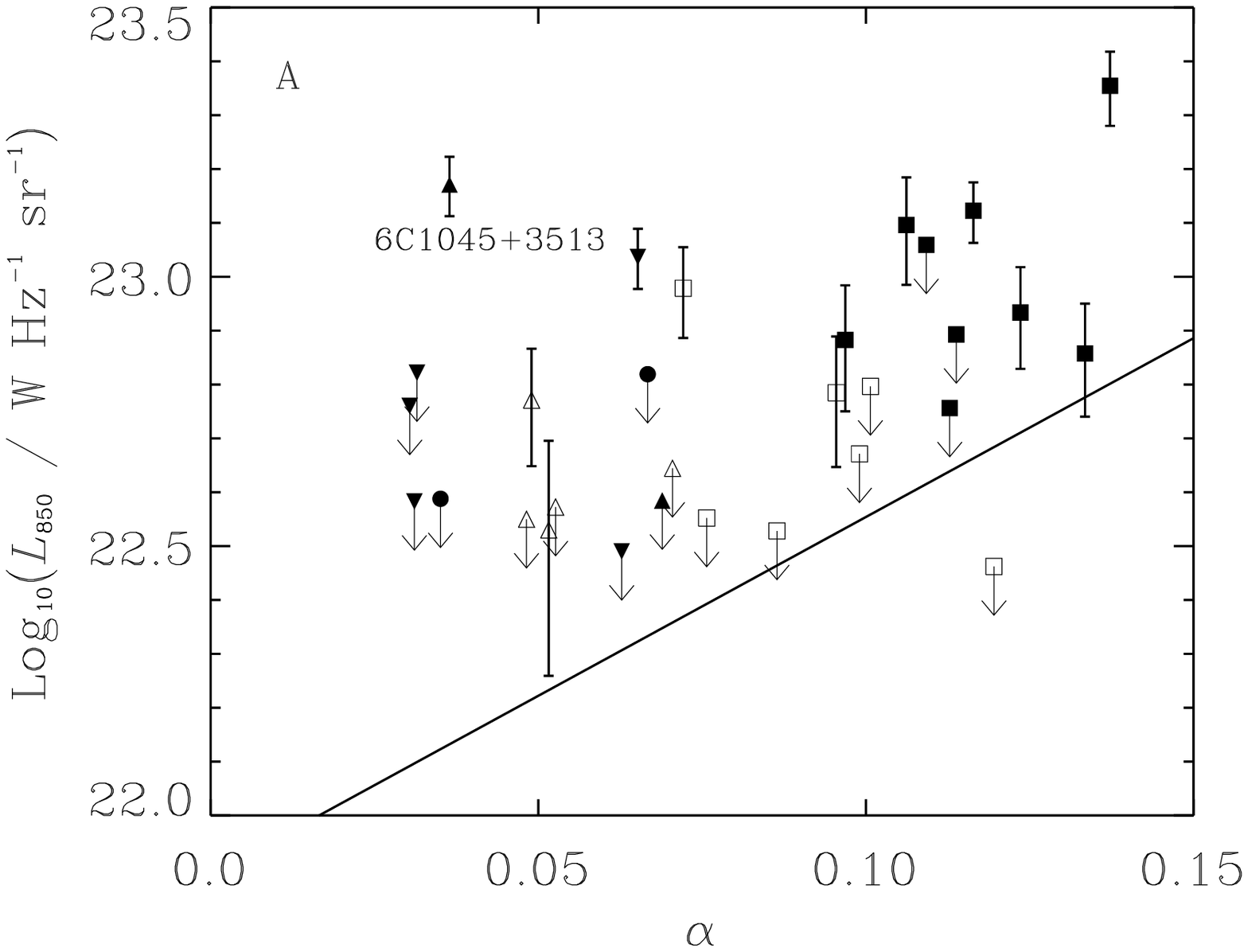}} 
\put(75,0){\includegraphics{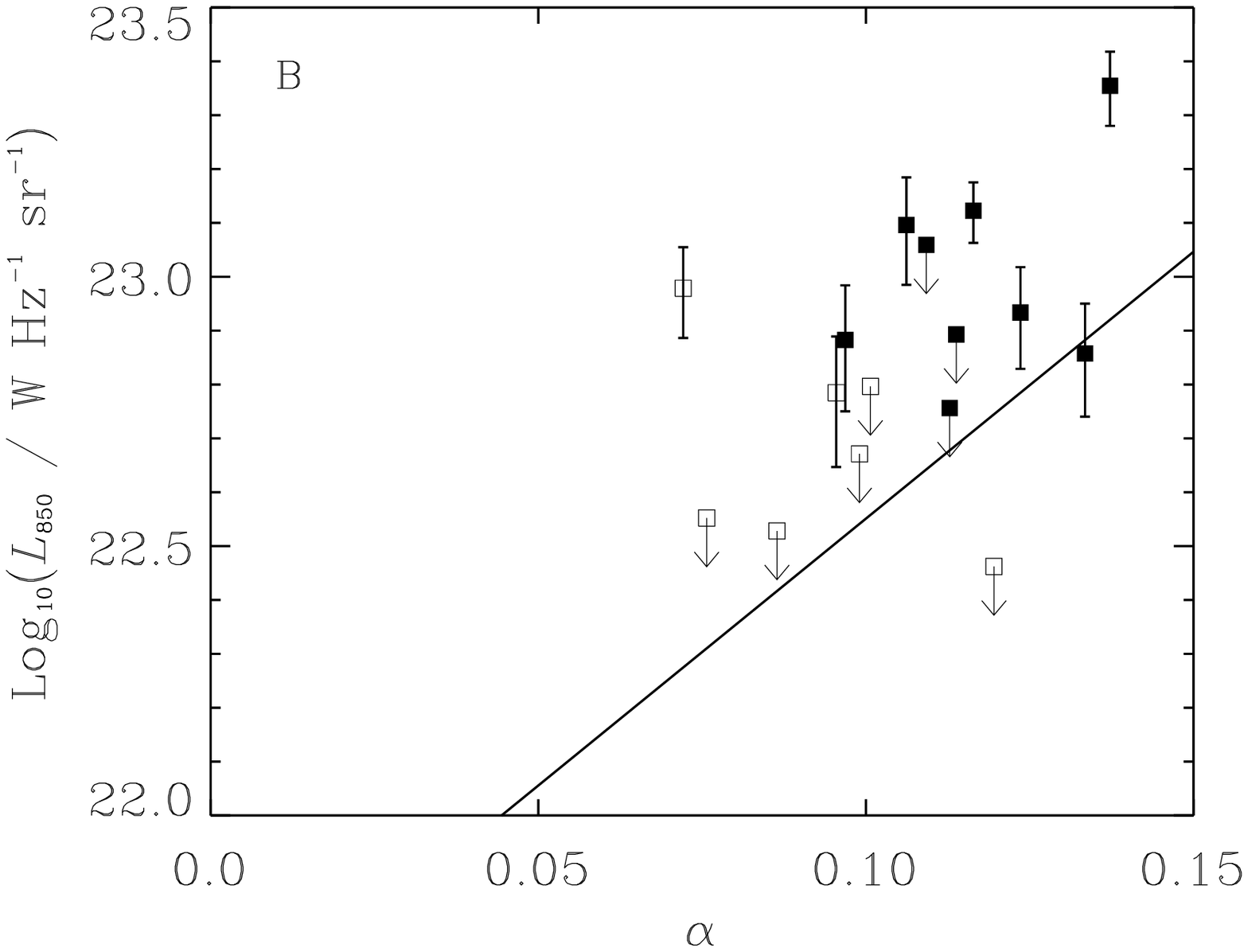}}
\end{picture}
\begin{picture}(150,60)
\put(40,0){\includegraphics{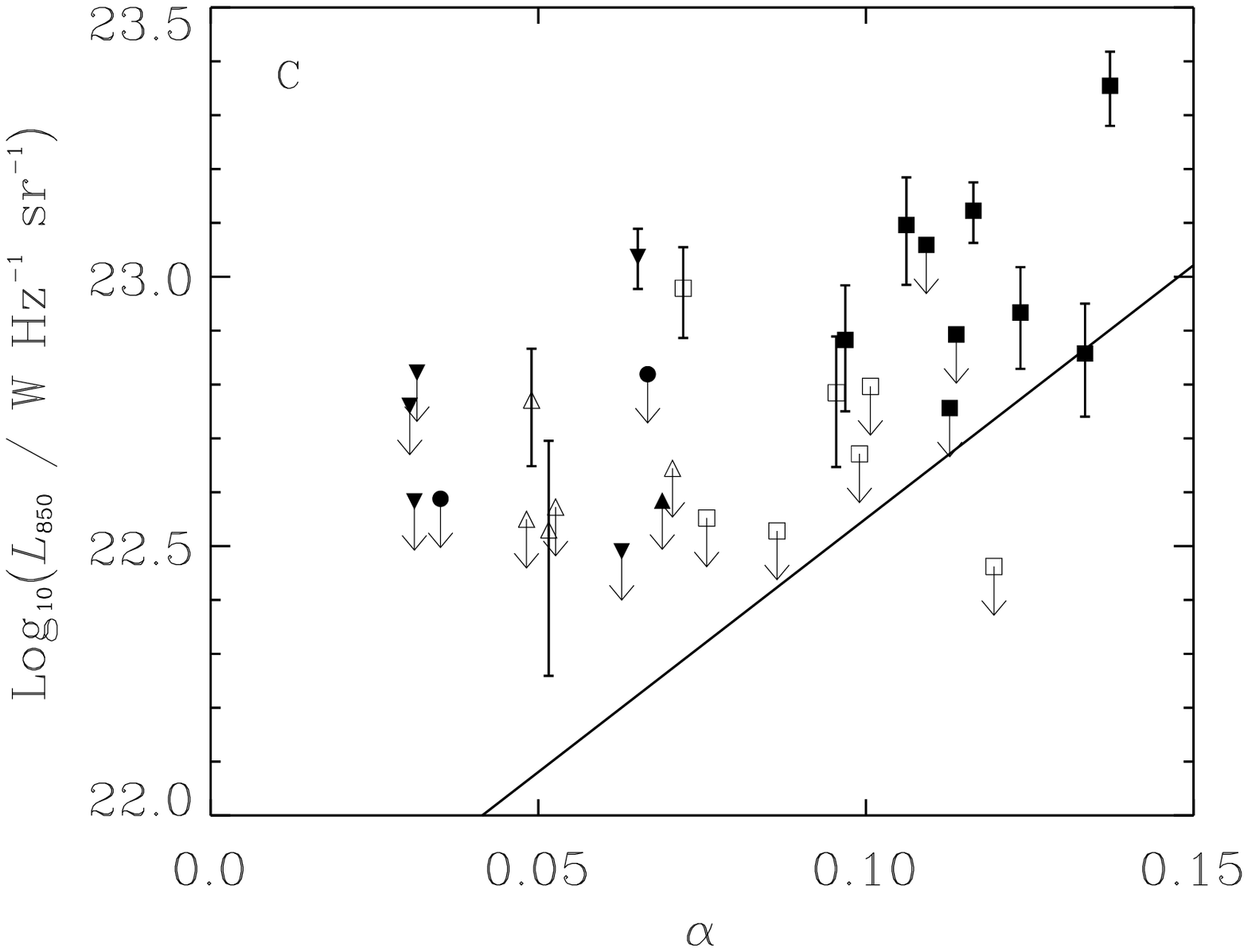}}
\end{picture}
\end{center}
\vspace{-1cm}
{\caption[junk]{\label{fig:alphaL3}{
Graphical representation of the results of the optimization procedure 
described in Sec.~\ref{sec:ch3compwill} and Table~\ref{tab:compabc}.
The lines showing the best-fitting values for the $L_{850} - \alpha$ correlation 
for datasets A, B and C, are shown alongside the relevant data 
(symbols as in Fig.~\ref{fig:pzch3}).}}}
\end{figure*}

It follows that the optimized gradient for the strength of the $L_{850}-\alpha$ correlation 
may not be high enough to reproduce the submillimetre difference seen by \citet{wsubmm}. However,
this gradient was derived based on quasar data alone, and we have seen from the error
bars given in Table \ref{tab:ch3mainres}, and the width of the marginalized 
probability distribution in Fig.~\ref{fig:Pslope2}, that there is a large uncertainty in these values. 
We therefore repeated the optimization procedure with three free parameters 
($\sigma_L$, $G$ and $C$) and for three datasets defined
as follows: (A) the eighteen quasars from Table~\ref{tab:ch3data} 
and the twelve radio galaxies from Table~\ref{tab:rg}; (B) 
the eleven quasars and the twelve radio galaxies from \citet{wsubmm}; and (C)
the same dataset as (A), but excluding one radio quasar, 6C~1045+3513,
the most prominent outlier in Fig.\ref{fig:alphaL3}.

\begin{table*}
\begin{center}
\begin{tabular}{lcl@{\hspace{1.0mm}}c@{\hspace{1.0mm}}cl@{\hspace{1.0mm}}c@{\hspace{1.0mm}}ccc}
\hline\hline
\multicolumn{1}{c}{Dataset}&$\sigma_L$&\multicolumn{3}{c}{$G$}&\multicolumn{3}{c}{$C$}&2SS: $P_{\rm 2RS}$&2RR: $P_{\rm 2RS}$\\
\hline
\hline
A&$0.475^{+0.149}_{-0.114}$&$6.64$&$\pm$&$7.90$&$21.89$&$\pm$&$0.72$&$0.018$&$0.989$\\
B&$0.393^{+0.169}_{-0.118}$&$9.91$&$\pm$&$(7.0)$&$21.56$&$\pm$&$(1.0)$&$0.034$&$0.898$\\
C&$0.438^{+0.145}_{-0.109}$&$9.41$&$\pm$&$14.66$&$21.61$&$\pm$&$1.39$&$0.015$&$1.022$\\
\hline\hline
\end{tabular}
\end{center}
\caption{\label{tab:compabc} The best-fitting parameters $\sigma_L$, $G$ and $C$
for datasets A, B and C (see Sec.~\ref{sec:ch3compwill} for a definition of
these datasets). 
`Formal' error bars on the parameters are 
obtained from the $\nabla \nabla \eta$ matrix in the normal manner (see GRW), and
the probability $P_{\rm 2RS}$ of each of the models with
respect to model 2RS is calculated using 
Equation 12 of GRW.
Brackets indicate that the error is an estimate, 
because formal errors could not be found for these parameters 
(see Sec.~\ref{sec:ch3compwill} for a discussion of how these were estimated).}
\end{table*}

In Table \ref{tab:compabc} we present the results of 
the optimization procedure, and the lines using these best-fitting gradients
and offsets are plotted in Fig.~\ref{fig:alphaL3}. We find that models 2RS, 2RR 
and 2SS all give very similar optimized parameters for
a given dataset (A, B or C), and that for all datasets 
models with a receding torus (either 2RR or 2RS) in the high-$\alpha$ population are favoured
strongly over the 2SS model. Formal
errors could not be found for the gradient $G$ and the offset $C$ for dataset B because
the range in $\alpha$ for this dataset
is so small that the likelihood surfaces are not well described by a 
multi-dimensional-Gaussian approximation. To find 
an error on $G$ (given in the brackets
in Table \ref{tab:compabc}), we fitted a Gaussian to the marginalized
probability distribution of the gradient (Fig.~\ref{fig:Pslope2}); the 
probability distributions for $G$ given the 2RR model and the various datasets are very similar. 
The error on $C$, given in the brackets in Table \ref{tab:compabc}, is a rough estimate 
based on the expectation that it is greater than that for dataset A, because of the small range
in $\alpha$, but not as great as that for dataset C, which arises because of the  
double-peaked shape of the probability distribution which is more pronounced than
that for dataset A (Fig.~\ref{fig:Pslope2}). These peculiar double-peaked shapes arise 
because of the large fraction  of upper limits in the datasets. Note, however, that for
all datasets, a gradient of zero is strongly disfavoured because the 
upper limits are concentrated at low values of $\alpha$. Note also, that the secondary
peaks at $S \sim 20$ are strongly disfavoured by the statistical detection of the
TOOT quasars in the submillimetre by \citet{rawlings}, so ignoring these, the peaks in the 
marginalized probability distributions all cluster in the range $6 \ltsimeq G \ltsimeq 10$.

We see in Fig.~\ref{fig:Pslope2} that the large uncertainty in the gradient $G$ is also 
compounded by the dramatic influence of just one outlying point (6C~1045+3513).
It is interesting that the peak of the probability distribution for $G$ shifts to significantly 
higher values for datasets B and C, in which this prominent outlier is excluded.
It is probable that the submillimetre emission for this object is not dominated by 
quasar-heated dust \citep{wsubmm}, but dominated instead by starburst-heated dust. The small size
of its radio jets indicate that it is a young radio source in which intense star-formation 
activity may be synchronized with the jet-triggering event. This hints that the correlation 
may be strongest for more typical radio sources in which quasar-heated dust may dominate even at the 
longer wavelengths probed in the submillimetre.

Simulations of the $S_{850} - L_{151}$ plane using the optimized parameters from dataset C 
were performed. Once more, as expected, the 2SS scheme gives identical 
values of $S_{850}$ ($\approx 3.40$ Jy) for radio quasars and radio galaxies.
For the other schemes, larger shifts are confirmed: an average observed flux $S_{850}$ of $4.65$ Jy for
radio quasars and only $2.09$ Jy for radio galaxies for 2RS; and $4.39$ Jy for radio quasars
and $2.20$ Jy for radio galaxies for 2RR. Moreover,
Fig.~\ref{fig:Pslope2} shows that it
is perfectly plausible that the true gradient is large enough that the difference in submillimetre 
luminosities of radio quasars and radio galaxies could easily be by a factor $\gtsimeq 2$. 
Accounting for
residual uncertainties in the corrections for synchrotron contamination \citep{wsubmm}, the factor 
$\sim 2-5$ difference in submillimetre luminosities seen by \citet{wsubmm} are now
in quantitative accord with both the 2RS and 2RR models.

\subsection{Comparison with `matched-pair' experiments}
\label{sec:ch3compmeis}
We consider in this section whether the GLFs described in Sec.~\ref{sec:ch3glf}
can reproduce the results of \citet{meisenheimer} and \citet{vanb}.

\citet{meisenheimer} observed ten pairs of 3CR radio galaxies and radio quasars with ISOPHOT on ISO
at infrared wavelengths between $5$ and $180\mu$m. The sources were matched in 
redshift and $178$-MHz luminosity $L_{178}$, but spanned a large
redshift range of $0.05 < z < 2.02$.
The average value of 100-micron flux density $S_{100}$ was $293$ mJy 
for quasars and $300$ mJy for radio galaxies, leading
to their conclusion that quasars and radio galaxies cannot be distinguished by their
observed mid- and far-infrared properties. However, if we use $\beta_{\rm d} \approx 2$ and
the values of dust temperatures found by fitting to the data in \citet{meisenheimer}
[if no values were given, a value of $40$ K was assumed], we find
that the average (rest-frame) 100-$\mu$m luminosity $L_{100}$ was $38.69$ for quasars and $38.36$ for 
radio galaxies, so that their radio quasars are an average of $\sim 2$-times more luminous than
radio galaxies at $100 ~ \mu$m. 

We simulated histograms of $\log_{10}L_{100}$ for the \citet{meisenheimer} 
dataset from the average of $30$ simulations using the optimized parameters
for dataset C (Sec.~\ref{sec:ch3compwill}). 
For 2SS, it is $38.57$ for radio quasars and $38.39$ for radio galaxies, with a shift of $0.18$
because the range of $z$ is sufficiently large that the low- and high-$\alpha$ 
populations become mixed, as was the case for the experiments of GRW.
For 2RS, the median $L_{100}$ is $38.75$ for quasars and $38.27$ for radio galaxies
giving a shift of $0.48$.  For 2RR, it is $38.75$ for quasars and $38.21$ for radio
galaxies, with a shift of $0.54$, so that for all models, radio quasars are expected to be
brighter than radio galaxies, in agreement with the data. 
To try and account for the effects of the selection of matched pairs, 
we used the GLFs to pick a value of $L_{100}$ (by extrapolating from the value of $L_{850}$ using our
canonical dust spectrum) from the actual redshift and $L_{178}$ values of each quasar and radio galaxy
in the \citet{meisenheimer} sample. We find that in only $\approx 60$ per cent of the pairs,
is the quasar brighter than the radio galaxy in $L_{100}$, based on one thousand 
simulations of picking the $10$ pairs. This shows that the
matched pairs method is not very effective at discovering quite large differences in the
properties of radio quasars and radio galaxies. \footnotemark

\footnotetext{
Note also that in one of the \citet{meisenheimer} matched pairs, 3C~325 (the
brighter far-IR source of its pair)
is taken to be a radio galaxy, whereas our new optical spectroscopic data (Appendix A) suggests
it is a reddened quasar.}

A  similar `matched pair' experiment was performed at lower redshift ($0.25 < z < 0.55$)
by \citet{vanb}.
They obtained ISOPHOT data for four pairs of 3CR radio galaxies and radio quasars, 
matched in $178$ MHz luminosity and redshift. They concluded that broad-lined objects
were brighter in the far-infrared than narrow-lined objects, since the average 90-$\mu$m luminosity
$L_{90}$ was $37.90$ for radio quasars and $\le 37.76$ for 
radio galaxies. Simulating their experiment using our GLFs we could reproduce 
this measured difference between radio quasars and radio galaxies for any of the unified
schemes. We note that the sample of \citet{vanb} contains such a small number of objects 
that it would be very difficult to reach significant conclusions from such an experiment. 

In conclusion, existing matched-pair experiments typically mix low- and high-$\alpha$ 
populations so all the GLFs predict a luminosity difference between radio quasars and radio
galaxies at $100 ~ \mu$m. We have seen that using small numbers of
matched pairs is not a particularly good way
of comparing unified schemes, but that the results from such methods are consistent 
with the other results described in this paper.

\section{Scaling relation between $L_{850}$ and $L_{\rm Bol}$}
\label{sec:scaling}

Here, we derive a rough scaling relation between $L_{850}$ and bolometric quasar luminosity
$L_{\rm Bol}$ for the high-$\alpha$ radio source population, argue that it may be applicable to 
quasars in general, and suggest how it might physically arise.

The form of the underlying $L_{850}$--$L_{\rm Bol}$ relation can be inferred
from the gradient $G$ (Table~\ref{tab:compabc}; Fig.~\ref{fig:Pslope2}). Using the value of $G$ derived from dataset C
(see Sec.~\ref{sec:ch3compwill}), we find that

\begin{equation}
\alpha \propto \frac{1}{9.41} \log_{10}L_{850}.
\end{equation}

From Equations 1 and 2 of GRW, we have
another expression for $\alpha$

\begin{equation}
\alpha \propto \frac{1}{s_1 \sqrt{2} \sqrt{n}} \log_{10}L_{151} + 
\frac{1}{s_2 \sqrt{2} \sqrt{n}} \log_{10}L_{\rm [OIII]}, 
\end{equation}

\noindent
in which $s_1 = 1.05$, $s_2 = 1.10$, and $n=302$.

\noindent 
From Equation 4 of GRW, we find that 

\begin{equation}
\log_{10}L_{\rm [OIII]} \propto 1.045 \log_{10}L_{151},
\end{equation}

\noindent so that if we assume that $L_{\rm Bol} \propto L_{\rm [OIII]}$, then

\begin{equation}
\label{eqn:scaling}
L_{850} \propto L_{\rm Bol}^{0.7 \pm 0.2},
\end{equation}

\noindent
where the $1\sigma$-error bar has been estimated 
from the probability distribution of the gradient (Fig.~\ref{fig:Pslope2}, ignoring the 
secondary peak) rather than from the value from the covariance matrix. Note that 
adoption of dataset A would be consistent with Eqn.~\ref{eqn:scaling},
but would favour gradients towards the lower end of the quoted error bar. Note also that values derived
from secondary peaks in the marginalized probability distributions of $G$ are inconsistent
with the constraints set from considering the sample consisting of only radio quasars
[see Fig.~\ref{fig:Pslope2} and the statistical detection of TOOT quasars by
\citet{rawlings}], and seem likely to be artefacts of the small ranges in 
$\alpha$ probed by submillimetre-detected objects in the relevant datasets. 
The scaling relation of Equation 9 is in good agreement with the results of Sec.~\ref{sec:PCA} in which 
it was found that 
the first principal component represented a linear (roughly equal weight) combination of
normalized versions of $\log_{10} L_{151}$, $\log_{10} L_{[OIII]}$ and $\log_{10} L_{850}$, 
meaning that the 
three variables scale-up with each other according to power laws with indices given roughly by 
the ratios of the associated data standard deviations, namely 
$\sim 1:1:0.6$ (assuming the real values of $\log_{10} L_{850}$ lie close to their
1-$\sigma$ upper limits).

The derived $L_{850}$--$L_{\rm Bol}$ scaling relation is similar to that inferred from the correlation between
$L_{850}$ and the optical luminosity of radio-quiet quasars: \citet{wrg}
found a factor $\sim 1.4$ reduction in $L_{850}$ per unit drop in 
absolute blue magnitude $M_{B}$, which corresponds to $L_{850} \propto L_{\rm Bol}^{0.4}$
if we assume that $L_{\rm Bol}$ scales linearly with the observed optical luminosity.
This suggests that the scaling relation derived here is not peculiar to radio-selected objects, but is a generic
feature of AGN.

There is one obvious way in which this scaling might arise. Consider models in which the 
observed submillimetre luminosity of powerful AGN comes from the dust in the obscuring torus
as a result of direct quasar heating. As suggested previously by \citet{sr}, a receding-torus-like scheme
means that the solid angle subtended by the torus scale with radius $r$ and
hence $L_{\rm Bol}$ as $h / r \propto L_{\rm Bol}^{-0.5}$
for constant torus half-height $h$. If the fraction of the radiation re-processed by the dust torus is 
simply proportional to this solid angle, then the implied scaling relation is 
$L_{850} \propto L_{\rm Bol}^{0.5}$. This is within the errors of the value 
observed (Eqn.~\ref{eqn:scaling}). Of course, 
more realistic receding torus models might feature distributions in $h$ and/or systematic variations
of $h$ with $L_{\rm Bol}$ \citep{sconf}, and hence predict slightly different scaling relations

\section{Conclusions}
\label{sec:conclusions}

In this paper, the generalized-luminosity-function (GLF) approach of
GRW has been extended to predict the number density of low-radio-frequency-selected
quasars and galaxies as a 
function of redshift, radio luminosity, emission-line luminosity and $850$-$\mu$m luminosity,
or observed $850$-$\mu$m flux density and sensitivity, using a new method which enables us to
use data with upper flux-density limits. The new free parameters in these extended GLFs were fixed 
by comparison with SCUBA ($850$-$\mu$m) data on high-luminosity, high-redshift radio sources.
Results from our GLF approach were found to
reproduce previous comparitive studies of the submillimetre and far-infrared luminosities
of radio quasars and radio galaxies \citep{wsubmm,meisenheimer,vanb}.

We find robust evidence that
the submillimetre luminosity of low-frequency-selected radio sources is
correlated with the GRW $\alpha$ parameter, and hence the bolometric quasar
luminosity $L_{\rm Bol}$ via a scaling relation $L_{850} \propto L_{\rm Bol}^{0.7 \pm 0.2}$.
This corresponds to a factor of $\approx 1.9 \pm 0.3$ reduction in $L_{850}$ per unit drop in 
absolute blue magnitude $M_B$ which is similar to the correlation tentatively
identified for the radio-quiet quasar population \citep{wrg}. The 
fact that the gradient of the scaling relation is less steep than a proportionality may give clues as to how 
$L_{850}$ arises, and is consistent with models in which quasar heating 
dominates, and in which dust in the receding torus intercepts and reprocesses a lower fraction 
of the bolometric luminosity into the mid- and far-infrared bands.

We also find robust evidence that a receding-torus-like scheme is strongly favoured 
(by a factor $\sim 40$ in probability over models without a receding torus) in the 
high-$\alpha$ population
of a two-population scheme for radio sources. In such schemes, radio quasars are brighter in
narrow emission lines and submillimetre luminosity (than radio galaxies matched in radio luminosity and
redshift) because they are biased towards the objects with the highest values within the 
spread of bolometric luminosity.

\section*{Acknowledgements}
JAG was supported by a graduate studentship from Oxford University and
Magdalen College. SR thanks the PPARC for a Senior Research Fellowship.
We thank Matt Jarvis and Chris Simpson for useful conversations, and the
anonymous referee for a very useful report.
This research has made use of the NASA/IPAC Extragalactic Database, which
is operated by the Jet Propulsion Laboratory, Caltech, under contract
with the National Aeronautics and Space Administration.

\appendix

\section{Corrected redshift and classification for 3C~325}

In this Appendix we present a new optical spectrum of 3C~325
which has allowed us to correct both the redshift and classification of this
object. We targeted this object because it 
is highly unusual amongst high-redshift ($z > 0.5$) radio galaxies in the 
3CR catalogue at it has been detected (using ISO) as a luminous
far-infrared source \citep{meisenheimer}. High-$z$ radio sources detected in the
far-infrared by IRAS and ISO are almost excusively
quasars, sometimes because the far-infrared
emission is from a beamed non-thermal component \citep{hbf}, but sometimes
because, as detailed in \citet{wsubmm} and this paper, radio quasars seem 
to be genuinely brighter (rest-frame) far-infrared sources than radio 
galaxies when matched in radio luminosity and redshift. Our previous
optical spectroscopy of another anomalous far-infrared-loud
3CR `galaxy', 3C~318, found that this object had an incorrect redshift
and classification in the literature, and that the object was
actually a $z= 1.574$ reddened quasar \citep{wrj}. Prior to this work, 
3C~325 was classified as an `N' galaxy with a redshift $z=0.86$ 
based on the identification of spectral features with 
moderate, possibly broad, Mg 2799 and weak [OII] 
3727 emission \citep{spinrad}.

\begin{figure*}
\begin{center}
\setlength{\unitlength}{1mm}
\begin{picture}(150,150)
\put(150,0){\includegraphics{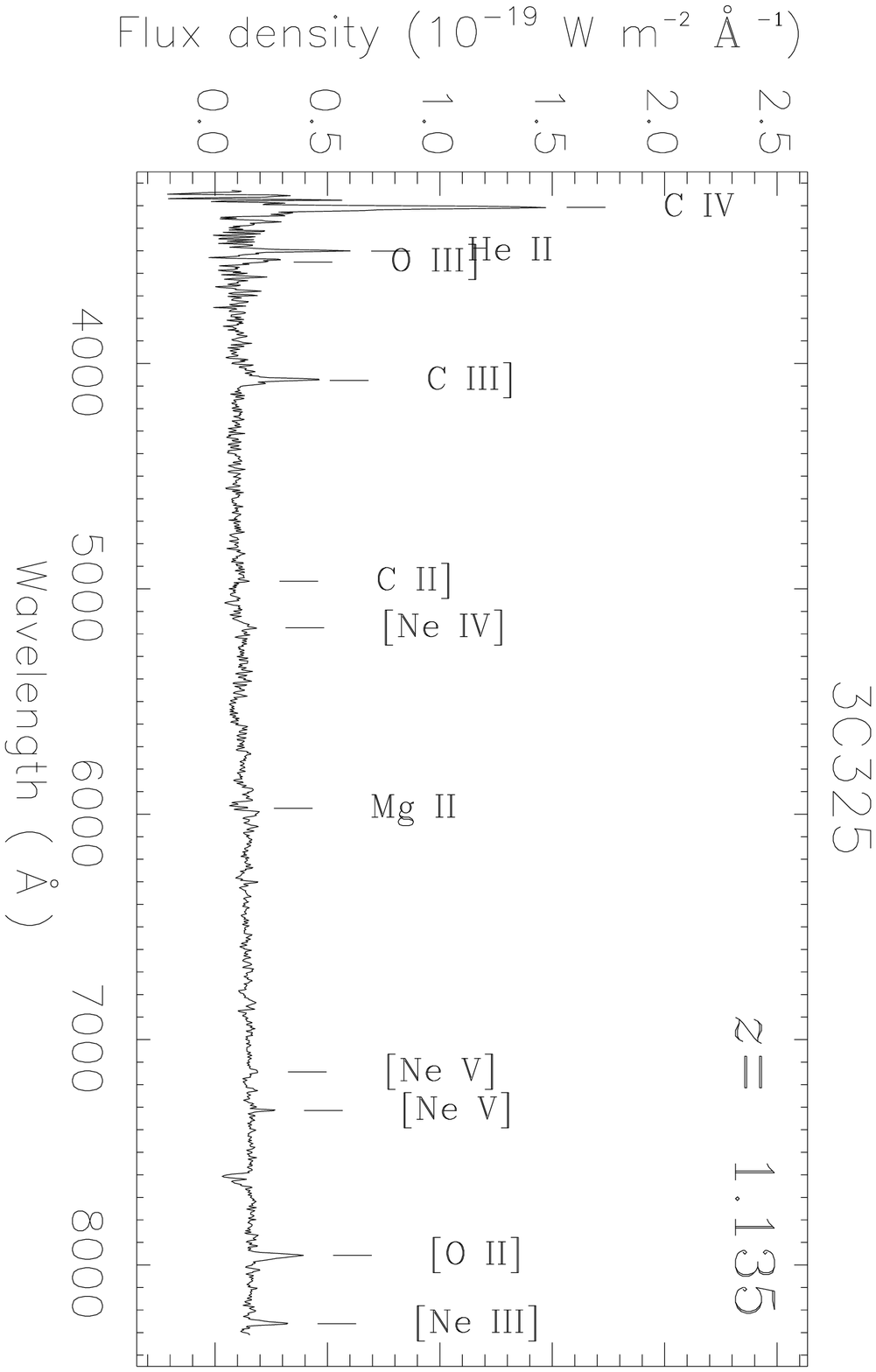}}
\end{picture}
\end{center}
{\caption[junk]{\label{fig:spec}
{
Optical spectrum of the reddened
quasar 3C~325, with the emission lines labelled.
}}}
\end{figure*}

3C~325 was observed with the dual-beam ISIS spectrograph on
the William Herschel Telescope on 22 April 2003; the 
R300B and R316R gratings were used. Conditions were
photometric with good ($\approx 0.8$ arcsec) seeing and a 1.5-arcsec slit 
was used. The telescope was pointed at the known optical identification
of 3C~325 (RA 15:49:58.22, DEC +62:41:22.4, J2000.0). 
The total integration time of the observation was 2700s, split into
three separate exposures on each arm. 
In Figure~\ref{fig:spec} we show the extracted optical spectrum
which was spatially unresolved in the 
two-dimensional spectrum. 
The object has many strong emission lines (see Table A1)
and therefore an unequivocal redshift of $z=1.135$; the previously reported 
redshift is incorrect, presumably because the [NeIV] 
line was confused with Mg II.

Although the cores of most of the emission lines are narrow there is a definite
broad ($\sim 3000 ~ \rm km ~ s^{-1}$) component underlying the 
C III] line, and probably a similar component associated with 
C IV. The Mg II line maybe dominated by a broad feature
but it is hard to be certain because there seems to be a lot of 
absorption to the blue side of the line. The
red continuum has a spectral index of $\approx 2.5$, so this object
seems to have all the characteristics of a reddened quasar.

3C~325 is a reddened quasar at $z=1.135$, and not an `N' galaxy at 
$z=0.86$ as previously reported \citep{spinrad}. It is interesting that 
so many of the far-infrared-loud
radio sources at high redshift appear
to be in the reddened quasar class, as we have discussed in
\cite{wsubmm}.

\begin{table*}
\begin{center}
\begin{tabular}{rllcr}
\hline
{Line} & {$\lambda_{\rm obs} / $} &{$z_{\rm em}$} & {FWHM / }
&{Flux}  \\
&{\AA} & & {km s$^{-1}$} & { $10^{-19}$ W m$^{-2}$} \\

\hline
C\,{\footnotesize IV} $1549$ & 3309 & 1.136 & 1300$\dag$ &   42.0 (20)   \\
He\,{\footnotesize II} $1640$ & 3501 & 1.135 & 750  &    8.9 (30)   \\
O\,{\footnotesize III}] $1663$& 3547 & 1.133 &   ?  &   2.0 (50)   \\
C\,{\footnotesize III}] $1909$& 4071 & 1.133 & 800$\dag$ &   9.0 (30)  \\
C\,{\footnotesize II}] $2326$ & 4966 & 1.135 &   ?  &   0.8 (60) \\

[Ne\,{\footnotesize IV}] $2423$ & 5174 & 1.135 &   ?  &   1.7 (50) \\

Mg\,{\footnotesize II} $2799$ & 5990 & 1.140$\ddag$ & 5000$\ddag$  & 4.0$\ddag$ (60) \\

[Ne\,{\footnotesize V}] $3346$ & 7142 & 1.134 & 500 & 0.7 (40) \\

[Ne\,{\footnotesize V}] $3426$ & 7313 & 1.135 & 400 & 1.2 (30) \\

[O\,{\footnotesize II}] $3727$ & 7961 & 1.136 & 900 & 8.0 (20) \\

[Ne\,{\footnotesize III}] $3869$ & 8258 & 1.134 & 600 & 3.8 (25) \\

\hline
\end{tabular}

{\caption[junk]{\label{tab:lines} Emission line data for 3C~325
from our optical spectrum. Emission line widths have been deconvolved
assuming that the source of optical emission was spatially unresolved. $\dag$ 
denotes that this refers to the core of a line which appears to have
broad wings. $\ddag$ denotes that values are very uncertain because
of absorption. 
The line fluxes have been measured by integrating along the slit direction 
and no corrections for slit losses have been attempted; estimated errors
are expressed, in brackets, as a percentage of the measured flux.
}}

\normalsize
\end{center}
\end{table*}


\label{lastpage}

\end{document}